\pdfoutput=1
\documentclass[aps, prb, twocolumn]{revtex4-2}

\usepackage{enumerate}
\usepackage{comment}
\usepackage{amsmath}
\usepackage{amssymb}
\usepackage{color}
\usepackage[dvipsnames]{xcolor}
\usepackage{tikz}
\usepackage{amsthm}
\usepackage{graphicx}
\usepackage{pict2e}
\usepackage{bbold}
\usepackage{empheq}
\usepackage{subfigure}
\usepackage{xfrac}
\usepackage{hyperref}
\usepackage{enumitem}
\usepackage{notes2bib}

\renewcommand{\vec}[1]{\mbox{\boldmath$#1$}}

\newcommand{\eqn}{Eq.~}

\newcommand{\fig}{Fig.~}

\newcommand{\reffs}{Refs.~}

\newcommand{\BLG}{bilayer graphene}

\newcommand{\EDITMH}[1]{{\color{black} #1}}
\newcommand{\EDITLS}[1]{{\color{black} #1}}

\begin{document}

\title{Gate-tunable regular and chaotic electron  dynamics in ballistic bilayer graphene 
cavities} 
 
\author{Lukas Seemann$^{1}$}
\author{Angelika Knothe$^{1,2}$}
\author{Martina Hentschel$^{1}$}
\affiliation{$^1$Institute of Physics, Technische Universit\"at Chemnitz, D-09107 Chemnitz, Germany}
\affiliation{$^2$Institut f\"ur Theoretische Physik, Universit\"at Regensburg, D-93040 Regensburg, Germany}
\date{\today}


\begin{abstract}
{
The dispersion of any given material is crucial for its charge carriers' dynamics.
For all-electronic, gate-defined cavities in gapped \BLG{}, we developed a trajectory-tracing algorithm   aware of the material's electronic properties and details of the confinement. We show how the anisotropic dispersion of \BLG{}  induces chaotic and regular dynamics depending on the gate voltage, \EDITMH{despite the high symmetry of the circular cavity}. Our results demonstrate the emergence of non-standard fermion optics solely due to anisotropic material characteristics.
}
\end{abstract}
 
\maketitle

\emph{Introduction}.
In two-dimensional materials, the reduction of dimensionality combined with the  material-dependent electronic properties often result in unusual charge carrier dynamics. In graphene-based systems, specifically, the remarkable sample quality 
due to low strain and charge disorder allowed the observation of  ballistic charge carrier dynamics over tens of $\mu m$ distances \cite{banszerusBallisticTransportExceeding2016, leeBallisticMinibandConduction2016, goldCoherentJettingGateDefined2021, berdyuginMinibandsTwistedBilayer2020}. Advanced gating techniques give  immense control over the charge carriers and the materials' electronic properties. This progress has recently led to the demonstration of electron confinement and control in a series of high-quality gate-defined quantum devices in \BLG{}, including quantum wires and  dots \cite{eichCoupledQuantumDots2018, overwegElectrostaticallyInducedQuantum2018, overwegTopologicallyNontrivialValley2018, banszerusSpinpolarizedCurrentsBilayer2019, banszerusGateDefinedElectronHole2018, leeTunableValleySplitting2020, goldCoherentJettingGateDefined2021, mollerProbingTwoElectronMultiplets2021, tongTunableValleySplitting2021, knotheQuartetStatesTwoelectron2020, laneSemimetallicFeaturesQuantum2019, garreisShellFillingTrigonal2021, knotheTunnelingTheoryBilayer2022}, Josephson junctions, interferometers, and SQUIDS \cite{iwakiriGateDefinedElectronInterferometer2022, rodan-legrainHighlyTunableJunctions2021, devriesGatedefinedJosephsonJunctions2021, portolesTunableMonolithicSQUID2022, dauberExploitingAharonovBohmOscillations2021, elahiDirectEvidenceKleinantiKlein2022}. 

Ballistic charge carrier dynamics in monolayer graphene cavities have been discussed previously in the context of Dirac electron optics in analogy with optical cavities. 
Early works on relativistic quantum chaos in graphene based on the conical Dirac dispersion and the chiral charge carriers discuss, e.g., Klein tunnelling and Veselago lens focussing, relativistic quantum scars, the interplay of chaos and spin  manifesting as spin-dependent, coexistent regular and chaotic scattering, and the  appearance of bound states  as resonances in the  conductance across integrable cavities \cite{cheianovFocusingElectronFlow2007, peterfalviElectronFlowCircular2009, xuRelativisticQuantumChaos2021, xuChaosDiracElectron2018, schneiderResonantScatteringGraphene2011, schneiderDensityStatesProbe2014, ponomarenkoChaoticDiracBilliard2008, huangPerspectivesRelativisticQuantum2020, heinlInterplayAharonovBohmBerry2013, hanDecaySemiclassicalMassless2018, geImagingQuantumInterference2021, brunGrapheneWhisperitronicsTransducing2022, miaoPhaseCoherentTransportGraphene2007, bardarsonElectrostaticConfinementElectrons2009, schrepferDiracFermionOptics2021, elahiDirectEvidenceKleinantiKlein2022, PhysRevB.103.L081111}.
\EDITMH{The correspondence of trajectory and wave calculation results for graphene systems was confirmed in Ref.~\onlinecite{schrepferDiracFermionOptics2021} and is the basis of the present work.}

In this Letter we present the  
\EDITMH{particle}
trajectory dynamics of ballistic charge carriers in circular, all-electronic gate-defined cavities in gapped \BLG{}, see \fig{}\ref{fig:1}. The unique material properties of \BLG{} induce  novel  trajectory dynamics yet distinct from the Dirac and the photonic case. The trigonally warped dispersion entails an anisotropic, valley-dependent velocity distribution of the charge carriers and unusual Snell and Fresnel laws at the cavity boundary (including anti-Klein-like tunnelling \cite{katsnelsonChiralTunnellingKlein2006, tudorovskiyChiralTunnelingSinglelayer2012, snymanBallisticTransmissionGraphene2007, duTuningAntiKleinKlein2018b, gradinarConductanceAnomalyLifshitz2012, nilssonTransmissionBiasedGraphene2007,  nakanishiRoleEvanescentWave2011, milovanovicBilayerGrapheneHall2013, parkBandGapTunedOscillatory2014, nakanishiTransmissionBoundaryMonolayer2010, barbierKronigPenneyModelBilayer2010, vanduppenFourbandTunnelingBilayer2013, sanudoStatisticalMagnitudesKlein2014, saleyKleinTunnelingTriple2022, heChiralTunnelingTwisted2013},  nonlinear relations between the angles of incidence, reflection, and transmission, and  total reflection due to momentum mismatch or the Fermi energy lying inside the band gap).

Unusual dynamics and Snell's laws due to anisoptric dispersions have been observed, e.g., for spin waves \cite{stigloherSnellLawSpin2016}, plasmons \cite{ahnTheoryAnisotropicPlasmons2021}, and in anisotropic photonic crystals \cite{alagappanSymmetryPropertiesTwodimensional2006, khromovaAnisotropicPhotonicCrystals2008}. In   \BLG{}, the anisotropic shape of the Fermi surface is enhanced by an interlayer asymmetry gap and depends on the charge carrier density \cite{peterfalviIntrabandElectronFocusing2012, goldCoherentJettingGateDefined2021}. Therefore, the carrier dynamics and scattering laws in \BLG{} can be tuned with external gates \cite{schrepferDiracFermionOptics2021}. Using a trajectory tracing algorithm that accounts for materials' characteristics, depending on the parameters chosen
we find:
\begin{figure}[t!]
    \centering
    \includegraphics[width=1\linewidth]{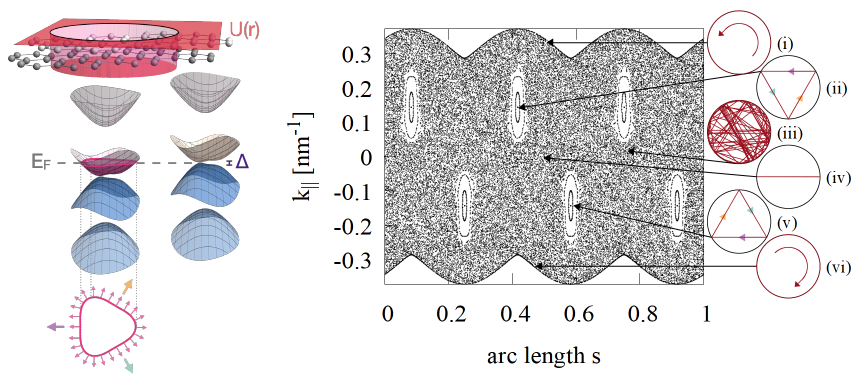}
    \caption{Left: circular cavity in bilayer graphene defined by a gate-defined potential $U(\mathbf{r})$. Right:  PSOS 
    for $\Delta=5 $ meV, $E_F=100$ meV, $U_0=100$ meV in the $K^+$ valley, where the Fermi energy is in the conduction band (gap) inside (outside) the cavity. We find chaotic dynamics (iii), stable triangular, periodic orbits (ii, v), unstable periodic orbits along the diameter (iv), and whispering-gallery-like orbits (i, vi). 
    }
    \label{fig:1}
\end{figure}
\begin{itemize}[align=right,itemindent=2em,labelsep=2pt,labelwidth=1em,leftmargin=0pt,nosep]
    \item  Chaotic charge carrier dynamics inside the cavity for large parts of parameter space even for  circular cavities  due to the broken rotational symmetry and anisotropy of \BLG{}'s dispersion;
\item Stable periodic triangular orbits induced by the threefold rotational symmetry per valley;
    \item 
    Whispering-gallery-like trajectories along the inner boundary stabilised 
    by total internal reflection before they eventually enter chaotic dynamics; 
\end{itemize} 



We evidence these different dynamical regimes by analysing the generalised Poincar\'e surface of section (PSOS)  
for the ballistic ray trajectories inside the cavity  (right panel of  \fig{}\ref{fig:1}) \EDITMH{ in terms of the normalised arc length, $s= \Tilde{s}/ 2 \pi R$ (with cavity radius $R$) along the cavity boundary and the conserved momentum, $k_{\parallel}$, at each reflection. \footnote{Note that $k_{\parallel}$ replaces the commonly used sine of the angle of incidence as momentum variable \cite{berryRegularityChaosClassical1981, meissSymplecticMapsVariational1992, meissCantoriStadiumBilliard1992} since it is the  conserved momentum in the present anisotropic situation, cf.~the Supplemental Material for more details.}} 
We generally find a mixed phase space with threefold symmetry per valley. The six fixed points correspond to two  stable periodic triangular orbits arising from the preferential directions of the charge carriers induced by the trigonally warped dispersion (\fig\ref{fig:1} left sketch, \EDITMH{bottom, color-coded}). The high symmetry of these triangular orbits, cf.~Fig.~\ref{fig:1}, lets them fulfill the conventional reflection law, unlike generic trajectories where incoming and outgoing angles differ.

\begin{figure*}[htb]
    \centering
    \includegraphics[width=1\linewidth]{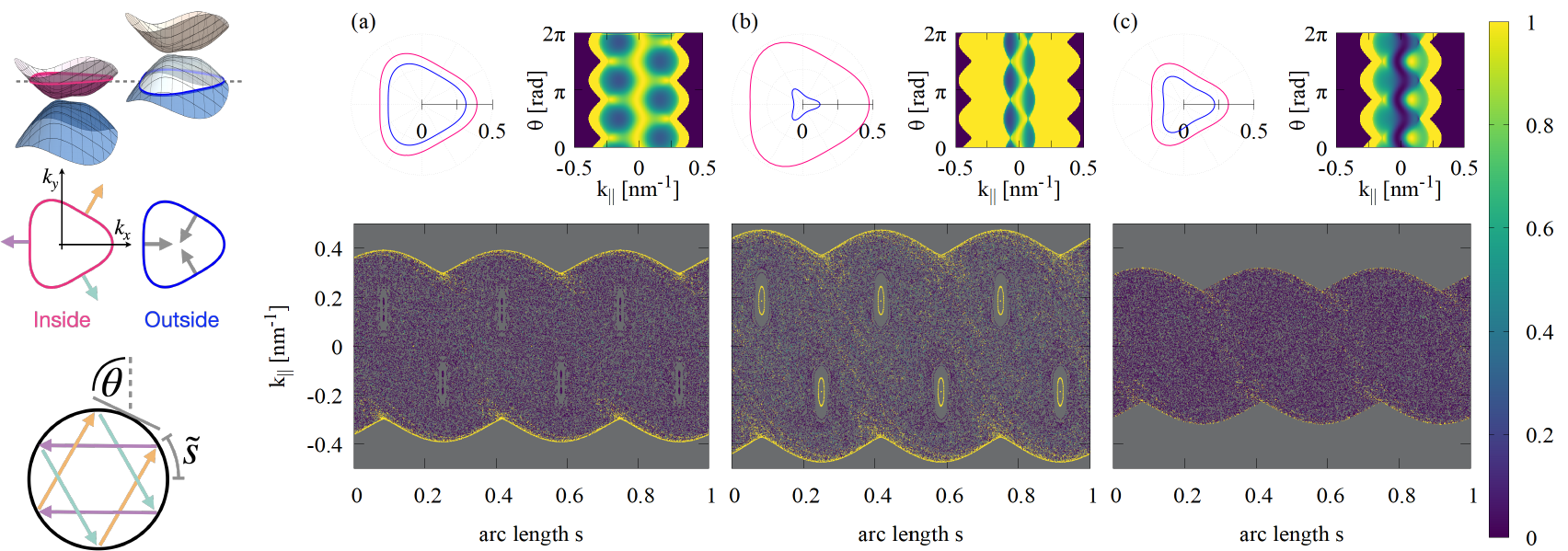}
    \caption{Dynamics of a np-type cavity for different values of the Fermi energy, the potential step, and the gap in the $K^+$ valley (a: $E_F= 100$ meV, $U_0= 180$ meV, $\Delta = 5$ meV; b: $E_F= 140$ meV, $U_0= 150$ meV, $\Delta = 5$ meV; c: $E_F= 70$ meV, $U_0= 120$ meV, $\Delta = 80$ meV). The geometry of the Fermi lines inside (magenta) and outside (blue) of the cavity, the local cavity coordinates, and the triangular trajectories are sketched on the left.  Top panels: 
    Inside/Outside Fermi lines (polar plots) and reflectivity R as a function of the orientation of the cavity boundary with respect to the lattice (density plots). The angle $\theta$ is defined in the  sketch, where $\theta=0$ corresponds to the armchair and $\theta=\pi/2$ to the zigzag orientation. Bottom panels: Intensity-weighted generalised PSOSs characterising the trajectory dynamics inside the cavity. Different Fermi lines inside and outside the cavity induce and stabilise different charge carrier dynamics. We use the same color scale for reflectivity in the top row, and intensity in the bottom row PSOS plots. 
    In the opposite valley, the Fermi lines are rotated by 180$^{\circ}$, inducing counterpropagating triangular orbits. We show the corresponding $K^-$ PSOSs in the Supplementary Material.}
    \label{fig:2}
\end{figure*}

This Letter is structured as follows. First, we derive  the laws for ballistic propagation and scattering at the cavity boundary in gapped \BLG{}. We then describe our particle tracing algorithm which takes into account these material specific propagation laws. Using this  algorithm, we study the ballistic charge carrier dynamics of \BLG{} either defined by a gapped region (cf. \fig \ref{fig:1}) or by an np-boundary (see \fig \ref{fig:2}). In conclusion, we discuss how our findings might be probed experimentally by transport or microscopy means.


\emph{Ballistic electron trajectory dynamics in \BLG{}}.
We  obtain the before-mentioned properties of ballistic charge carrier trajectory dynamics in \BLG{} cavities by performing material characteristics-aware ray tracing simulations. Our  ray-tracing algorithm 
takes into account the low-energy electronic structure of \BLG{} in the two inequivalent valleys $K^{\xi}$ ($\xi=\pm 1$) as described by the four-band Hamiltonian \cite{mccannLandauLevelDegeneracyQuantum2006, mccannLowEnergyElectronic2007, mccannElectronicPropertiesBilayer2013},
\begin{equation}
 H^{\xi}_{BLG}=\xi
\setlength{\arraycolsep}{-0pt} \begin{pmatrix} 
\xi U-\frac{1}{2}\Delta  & v_3\pi & v_4 \pi^{\dagger} &v \pi^{\dagger}\\
 v_3 \pi^{\dagger}&\xi U+\frac{1}{2}\Delta & v\pi &v_4 \pi\\
 v_4 \pi & v\pi^{\dagger} & \xi U+\frac{1}{2}\Delta  & \xi \gamma_1\\
 v\pi & v_4 \pi^{\dagger} & \xi \gamma_1 &\xi U-\frac{1}{2}\Delta 
\end{pmatrix}.
\label{eqn:H}
\end{equation}

Equation \eqref{eqn:H} is written in the sublattice basis $(A,B^{\prime},A^{\prime}, B)$ in the valley $K^{+}$ and $(B^{\prime},A, B,A^{\prime})$ in the valley $K^{-}$ 
with $\pi=p_x+i p_y$, $\pi^{\dagger}=p_x-i p_y$. Here $\Delta$ is an interlayer asymmetry gap, and $U(\mathbf{r})=U_0\mathcal{H}(\mathbf{r})$ is the potential step of height $U_0$ locally defining the cavity boundary (where $\mathcal{H}$ denotes the Heaviside step function). 
We include all couplings relevant at the low-energy scales under consideration, i.e., $v\approx 1.02 \cdot 10^6$ m/s, $v_3 \approx 0.12\cdot v$, $v_4\approx 0.37 \cdot v_3$, and $\gamma_1\approx 0.38$ eV \cite{kuzmenkoDeterminationGatetunableBand2009}.

From \eqn \eqref{eqn:H} follows that the electronic dispersion is anisotropic with trigonally warped iso-energy Fermi lines in each valley as shown in \fig\ref{fig:1}. This anisotropy of the dispersion entails an anisotropic velocity distribution of the charge carriers' real-space trajectories \cite{goldCoherentJettingGateDefined2021, peterfalviIntrabandElectronFocusing2012, kraftAnomalousCyclotronMotion2020}. Furthermore, the anisotropy translates into the laws for reflection and transmission  of charge carriers scattering off a potential boundary. We derive the reflectivity, $R$, and transmittivity, $T=1-R$, by wave matching across the sharp \footnote{Generalizations to smooth confinement potentials have been discussed, e.g., in \reffs \onlinecite{cheianovSelectiveTransmissionDirac2006, peterfalviIntrabandElectronFocusing2012, schrepferDiracFermionOptics2021}. See the conclusion for further discussion.} potential step at the cavity boundary \cite{katsnelsonChiralTunnellingKlein2006, tudorovskiyChiralTunnelingSinglelayer2012, snymanBallisticTransmissionGraphene2007, nilssonTransmissionBiasedGraphene2007,  nakanishiRoleEvanescentWave2011, milovanovicBilayerGrapheneHall2013, parkBandGapTunedOscillatory2014, nakanishiTransmissionBoundaryMonolayer2010, barbierKronigPenneyModelBilayer2010, vanduppenFourbandTunnelingBilayer2013, sanudoStatisticalMagnitudesKlein2014, saleyKleinTunnelingTriple2022, heChiralTunnelingTwisted2013}. Matching the eigenstates of \eqn \eqref{eqn:H} inside ($U\equiv0$) and outside ($U\equiv U_0$) of the cavity, cf.~\fig\ref{fig:1}, we determine the reflected and transmitted part of the wave, $\Psi_{r}$ and $\Psi_{t}$ and the corresponding reflectivity,
\begin{equation}
R(\theta)=\frac{\langle \Psi_{r}| v_{\perp}(\theta)|\Psi_{r}\rangle}{\langle \Psi_{in}| v_{\perp}(\theta)|\Psi_{in}\rangle},    
\end{equation}
where $\Psi_{in}$ is the amplitude of the incoming wave. 
Further, $v_{\perp}(\theta)={\partial  H^{\xi}_{BLG}}/{\partial p_{\perp}(\theta)}$ is the velocity component locally perpendicular to the cavity boundary in a coordinate system rotated by $\theta$ for each point around the cavity as sketched in \fig \ref{fig:2}. 
We provide details of the wave matching calculations in the Supplementary Material.

Note that as a consequence of the trigonally warped Fermi lines the reflection and transmission laws distinguish between the two valleys and depend on the orientation of the boundary with respect to the \BLG{} lattice. 
Further, the reflection and transmission coefficients, $R$ and  $T$, depend on the  gap, $\Delta$, the Fermi energy, $E_F$, and the height of the potential step, $U$, as we illustrate with the examples in \fig \ref{fig:2} top row. Adjusting these parameters hence allows for gate-defined engineering of the charge carriers' velocity distribution and Fresnel laws, and accordingly, the cavity dynamics. 

Our numerical ray-tracing algorithm generating the ballistic  \EDITMH{particle} trajectories of a \BLG{} billiard works as follows: For a given value of $\Delta, E_F$ and $U_0$, we determine the initial velocity distribution of the injected charge carriers according to the Fermi lines we obtain from diagonalising \eqn \eqref{eqn:H} in each valley \footnote{We neglect intervalley scattering and treat the two valleys separately. This is justified due to the high quality of current \BLG{} samples not hosting any atomic-scale defects or abrupt lattice termination \cite{goldCoherentJettingGateDefined2021}}. We start \EDITLS{20} trajectories at \EDITLS{36 points} on the cavity boundary according to this initial velocity distribution. Then, for every 
scattering 
at the cavity boundary, we apply the Fresnel laws for reflection and transmission prescribed by the system parameters and the relative orientation $\theta$ of the boundary at the point of intersection\EDITMH{, see the left of Fig.~\ref{fig:2}. The momentum component locally tangent to the boundary, $k_{\parallel}$, is conserved in each scattering.} 
The velocity orientation of the new ray trajectory after scattering is again re-calculated from the respective Fermi lines on either side of the boundary. \EDITLS{We trace each ray trajectory and  its intensity for 100 reflections.}

One may distinguish cases where the Fermi energy lies within the gap outside of the cavity (forcing total reflection at every \EDITMH{reflection point}) 
or np-junction boundaries (allowing for scattering into the hole band outside the cavity, see \fig{}\ref{fig:2} left sketch). While the former case amplifies the dynamics inside the cavity by preventing charge carriers from escaping (see \fig \ref{fig:1}), the latter allows to study far-field emission and offers the additional tuning nob of mismatching the Fermi contours inside and outside of the cavity boundary, cf.~\fig \ref{fig:2}.

In \fig \ref{fig:2}, we provide representative examples for distinct carrier dynamics obtained from our particle-tracing algorithm for np-type cavities with different values of the potential step, the Fermi energy, and the gap. Generally, a strong mismatch of the inside and outside Fermi contours \EDITMH{favors} 
total reflection whenever an electron state with a large conserved momentum cannot find a partner in the hole band. This tunability of total reflection allows to selectively capture certain trajectories inside the cavity. For example, in  \fig \ref{fig:2}(a),
where the Fermi surface have almost equal size and shape, total reflection due to momentum mismatch only occurs for large momenta, corresponding to whispering-gallery-like boundary modes (regions of high intensity only at the edges of the PSOS). However, if the outside Fermi surface is much smaller, as in \fig \ref{fig:2}(b), the regime of total reflection extends to smaller momenta (inner regions of the PSOS), also stabilizing the triangular orbits \EDITMH{that now keep high intensities upon propagation}. 

Figure \ref{fig:2}(c) exemplifies that for sufficiently small gaps and Fermi energies the Fermi line becomes concave.  Curving the legs of the triangular lines inwards entails a bifurcation of the charge carriers' velocity distribution and destroys the clear preferred directions previously prescribed by the triangle's flat legs  \cite{goldCoherentJettingGateDefined2021}. As a consequence, in the regime of concave Fermi lines inside the cavity 
the triangular \EDITMH{islands} 
are destroyed and the periodic orbits 
vanish.

\begin{figure}[htb]
    \centering
    \includegraphics[width=\linewidth]{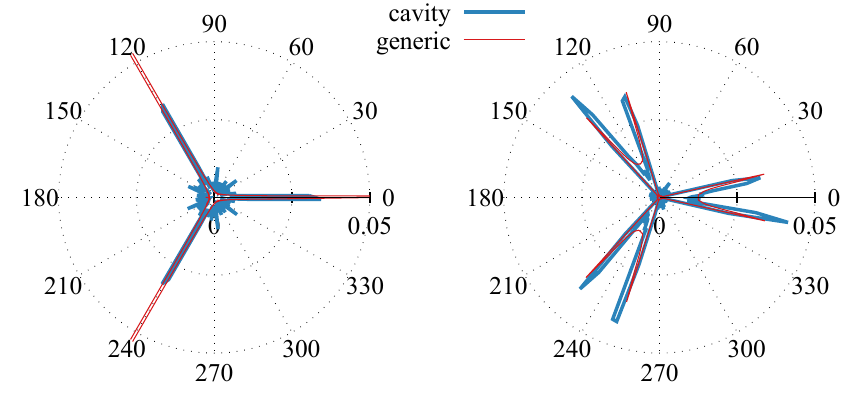}
    \caption{Far field emitted by a np-type cavity for the case of a convex (left) and concave (right) Fermi line in the p-doped outside region. We compare emission from the cavity (blue) to the emission from a generic, isotropic source (red) in BLG.}
    \label{fig:3}
\end{figure}

The bifurcation of the charge carriers' velocity distribution also manifests in the far field characteristics of the np-type cavities. We present the emitted far field in \fig \ref{fig:3} for examples where the Fermi lines outside the cavity are convex and concave. We note that the far-field is similar to that of a generic, isotropic source where all momentum states are populated equally, with the charge carrier dynamics being primarily prescribed by the velocity distribution following the outer Fermi line \cite{goldCoherentJettingGateDefined2021}. 

\emph{Conclusion and relation to experiment.}
We presented the ray trajectory dynamics of ballistic charge carriers in a circular gate-defined \BLG{} cavity. As a consequence of the anisotropic, trigonally warped dispersion, we find chaotic regimes, periodic triangular orbits, and 
\EDITMH{whispering-gallery-type} 
trajectories. \EDITMH{The charge carrier dynamics and the corresponding PSOS depend on system parameters which can be adjusted} 
by external gates. \EDITMH{Note that this plethora of structure here emerges in the PSOS of a circular billiard and does not require any deformation or anisotropy of the cavity shape \cite{berryRegularityChaosClassical1981, meissCantoriStadiumBilliard1992, xiaoAsymmetricResonantCavities2010, wiersigCombiningDirectionalLight2008}. The \BLG{} billiard hence constitutes an example how chaos is induced by material properties.}

In the present work, we  focused on the simple cases of a circular cavity with sharp boundaries, and a singly connected \BLG{} Fermi contour. Both the shape of the cavity boundary and the Fermi contour allow for additional tuning of the carrier dynamics. One may consider, e.g., deformed cavities \cite{berryRegularityChaosClassical1981, meissCantoriStadiumBilliard1992, wiersigCombiningDirectionalLight2008, xiaoAsymmetricResonantCavities2010}, smooth boundaries \cite{cheianovSelectiveTransmissionDirac2006, peterfalviIntrabandElectronFocusing2012, schrepferDiracFermionOptics2021},  and  Fermi lines below the \BLG{} Lifshitz transition which can be induced by external gates or strain \cite{varletTunableFermiSurface2015, varletAnomalousSequenceQuantum2014,gradinarConductanceAnomalyLifshitz2012, moulsdaleEngineeringTopologicalMagnetic2020}. Since the main ingredient of our analysis is the anisotropic dispersion and resulting velocity distribution, we anticipate the main features observed here, e.g.~, regular and chaotic charge carrier dynamics induced by the material dispersion, to manifest also in more complex setups. Moreover, one may perform equivalent considerations for other 2D materials with non-circular dispersions and corresponding anisotropies \cite{kraftAnomalousCyclotronMotion2020, leeBallisticMinibandConduction2016, berdyuginMinibandsTwistedBilayer2020}. Such further studies are straightforward generalisations of our work and will be discussed elsewhere.

Scanning gate microscopy has been used in the past to visualise electron trajectories in graphene and electronic cavities \cite{morikawaImagingBallisticCarrier2015, bhandariImagingFlowHoles2020, steinacherScanningGateExperiments2018, goldCoherentJettingGateDefined2021, chenScanningGateMicroscopy2022, goldImagingSignaturesLocal2021, fratusSignaturesFoldedBranches2021}. We anticipate that chaotic and periodic charge carrier trajectories would further be distinguishable in transport: Depending on the position of contacts around the cavity \footnote{Here, we do not take into account the possible distribution of states in an actual channel attached to the cavity, as, e.g., in \onlinecite{goldCoherentJettingGateDefined2021}. Doing so is a straightforward extension of our work to describe any actual experimental setup.}, one would record a different signal from a uniformly chaotic background than from trajectories focused into a detector by a periodic orbit. In a setup rotating contacts around the cavity, the number and position of any regions of enhanced conductance would allow conclusions about the Fermi surface's symmetry, topology, and orientation. Since the two valleys of \BLG{} couple to a magnetic field with the opposite sign \cite{knotheInfluenceMinivalleysBerry2018, leeTunableValleySplitting2020, tongTunableValleySplitting2021}, a weak out-of-plane magnetic field may be used to single out trajectories in different valleys. 



Overall, we have demonstrated how regular and chaotic charge carrier dynamics may arise due to inherent anisotropies and non-linearities of the material properties. In \BLG{}, the unique tunability of such material characteristics and the cavity confinement by external gates holds great potential for novel types of non-linear ballistic charge carrier dynamics.







\emph{Acknowledgements.} We thank Carolin Gold, Christoph Stampfer, Luca Banszerus, Caio Lewenkopf, Thomas Lane, Rafael A.~Molina, Yuriko Baba, Klaus Richter, Christoph Schulz, Ming-Hao Liu, and Sybille Gemming for valuable discussions.

\bibliography{BLGCavities.bib}

\begin{thebibliography}{82}%
\makeatletter
\providecommand \@ifxundefined [1]{%
 \@ifx{#1\undefined}
}%
\providecommand \@ifnum [1]{%
 \ifnum #1\expandafter \@firstoftwo
 \else \expandafter \@secondoftwo
 \fi
}%
\providecommand \@ifx [1]{%
 \ifx #1\expandafter \@firstoftwo
 \else \expandafter \@secondoftwo
 \fi
}%
\providecommand \natexlab [1]{#1}%
\providecommand \enquote  [1]{``#1''}%
\providecommand \bibnamefont  [1]{#1}%
\providecommand \bibfnamefont [1]{#1}%
\providecommand \citenamefont [1]{#1}%
\providecommand \href@noop [0]{\@secondoftwo}%
\providecommand \href [0]{\begingroup \@sanitize@url \@href}%
\providecommand \@href[1]{\@@startlink{#1}\@@href}%
\providecommand \@@href[1]{\endgroup#1\@@endlink}%
\providecommand \@sanitize@url [0]{\catcode `\\12\catcode `\$12\catcode
  `\&12\catcode `\#12\catcode `\^12\catcode `\_12\catcode `\%12\relax}%
\providecommand \@@startlink[1]{}%
\providecommand \@@endlink[0]{}%
\providecommand \url  [0]{\begingroup\@sanitize@url \@url }%
\providecommand \@url [1]{\endgroup\@href {#1}{\urlprefix }}%
\providecommand \urlprefix  [0]{URL }%
\providecommand \Eprint [0]{\href }%
\providecommand \doibase [0]{https://doi.org/}%
\providecommand \selectlanguage [0]{\@gobble}%
\providecommand \bibinfo  [0]{\@secondoftwo}%
\providecommand \bibfield  [0]{\@secondoftwo}%
\providecommand \translation [1]{[#1]}%
\providecommand \BibitemOpen [0]{}%
\providecommand \bibitemStop [0]{}%
\providecommand \bibitemNoStop [0]{.\EOS\space}%
\providecommand \EOS [0]{\spacefactor3000\relax}%
\providecommand \BibitemShut  [1]{\csname bibitem#1\endcsname}%
\let\auto@bib@innerbib\@empty
\bibitem [{\citenamefont {Banszerus}\ \emph {et~al.}(2016)\citenamefont
  {Banszerus}, \citenamefont {Schmitz}, \citenamefont {Engels}, \citenamefont
  {Goldsche}, \citenamefont {Watanabe}, \citenamefont {Taniguchi},
  \citenamefont {Beschoten},\ and\ \citenamefont
  {Stampfer}}]{banszerusBallisticTransportExceeding2016}%
  \BibitemOpen
  \bibfield  {author} {\bibinfo {author} {\bibfnamefont {L.}~\bibnamefont
  {Banszerus}}, \bibinfo {author} {\bibfnamefont {M.}~\bibnamefont {Schmitz}},
  \bibinfo {author} {\bibfnamefont {S.}~\bibnamefont {Engels}}, \bibinfo
  {author} {\bibfnamefont {M.}~\bibnamefont {Goldsche}}, \bibinfo {author}
  {\bibfnamefont {K.}~\bibnamefont {Watanabe}}, \bibinfo {author}
  {\bibfnamefont {T.}~\bibnamefont {Taniguchi}}, \bibinfo {author}
  {\bibfnamefont {B.}~\bibnamefont {Beschoten}},\ and\ \bibinfo {author}
  {\bibfnamefont {C.}~\bibnamefont {Stampfer}},\ }\bibfield  {title} {\bibinfo
  {title} {Ballistic {{Transport Exceeding}} 28 {$M$}m in {{CVD Grown
  Graphene}}},\ }\href {https://doi.org/10.1021/acs.nanolett.5b04840}
  {\bibfield  {journal} {\bibinfo  {journal} {Nano Letters}\ }\textbf {\bibinfo
  {volume} {16}},\ \bibinfo {pages} {1387} (\bibinfo {year}
  {2016})}\BibitemShut {NoStop}%
\bibitem [{\citenamefont {Lee}\ \emph {et~al.}(2016)\citenamefont {Lee},
  \citenamefont {Wallbank}, \citenamefont {Gallagher}, \citenamefont
  {Watanabe}, \citenamefont {Taniguchi}, \citenamefont {Fal'ko},\ and\
  \citenamefont {{Goldhaber-Gordon}}}]{leeBallisticMinibandConduction2016}%
  \BibitemOpen
  \bibfield  {author} {\bibinfo {author} {\bibfnamefont {M.}~\bibnamefont
  {Lee}}, \bibinfo {author} {\bibfnamefont {J.~R.}\ \bibnamefont {Wallbank}},
  \bibinfo {author} {\bibfnamefont {P.}~\bibnamefont {Gallagher}}, \bibinfo
  {author} {\bibfnamefont {K.}~\bibnamefont {Watanabe}}, \bibinfo {author}
  {\bibfnamefont {T.}~\bibnamefont {Taniguchi}}, \bibinfo {author}
  {\bibfnamefont {V.~I.}\ \bibnamefont {Fal'ko}},\ and\ \bibinfo {author}
  {\bibfnamefont {D.}~\bibnamefont {{Goldhaber-Gordon}}},\ }\bibfield  {title}
  {\bibinfo {title} {Ballistic miniband conduction in a graphene
  superlattice},\ }\href {https://doi.org/10.1126/science.aaf1095} {\bibfield
  {journal} {\bibinfo  {journal} {Science}\ }\textbf {\bibinfo {volume}
  {353}},\ \bibinfo {pages} {1526} (\bibinfo {year} {2016})}\BibitemShut
  {NoStop}%
\bibitem [{\citenamefont {Gold}\ \emph
  {et~al.}(2021{\natexlab{a}})\citenamefont {Gold}, \citenamefont {Knothe},
  \citenamefont {Kurzmann}, \citenamefont {{Garcia-Ruiz}}, \citenamefont
  {Watanabe}, \citenamefont {Taniguchi}, \citenamefont {Fal'ko}, \citenamefont
  {Ensslin},\ and\ \citenamefont {Ihn}}]{goldCoherentJettingGateDefined2021}%
  \BibitemOpen
  \bibfield  {author} {\bibinfo {author} {\bibfnamefont {C.}~\bibnamefont
  {Gold}}, \bibinfo {author} {\bibfnamefont {A.}~\bibnamefont {Knothe}},
  \bibinfo {author} {\bibfnamefont {A.}~\bibnamefont {Kurzmann}}, \bibinfo
  {author} {\bibfnamefont {A.}~\bibnamefont {{Garcia-Ruiz}}}, \bibinfo {author}
  {\bibfnamefont {K.}~\bibnamefont {Watanabe}}, \bibinfo {author}
  {\bibfnamefont {T.}~\bibnamefont {Taniguchi}}, \bibinfo {author}
  {\bibfnamefont {V.}~\bibnamefont {Fal'ko}}, \bibinfo {author} {\bibfnamefont
  {K.}~\bibnamefont {Ensslin}},\ and\ \bibinfo {author} {\bibfnamefont
  {T.}~\bibnamefont {Ihn}},\ }\bibfield  {title} {\bibinfo {title} {Coherent
  {{Jetting}} from a {{Gate-Defined Channel}} in {{Bilayer Graphene}}},\ }\href
  {https://doi.org/10.1103/PhysRevLett.127.046801} {\bibfield  {journal}
  {\bibinfo  {journal} {Physical Review Letters}\ }\textbf {\bibinfo {volume}
  {127}},\ \bibinfo {pages} {046801} (\bibinfo {year}
  {2021}{\natexlab{a}})}\BibitemShut {NoStop}%
\bibitem [{\citenamefont {Berdyugin}\ \emph {et~al.}(2020)\citenamefont
  {Berdyugin}, \citenamefont {Tsim}, \citenamefont {Kumaravadivel},
  \citenamefont {Xu}, \citenamefont {Ceferino}, \citenamefont {Knothe},
  \citenamefont {Kumar}, \citenamefont {Taniguchi}, \citenamefont {Watanabe},
  \citenamefont {Geim}, \citenamefont {Grigorieva},\ and\ \citenamefont
  {Fal'ko}}]{berdyuginMinibandsTwistedBilayer2020}%
  \BibitemOpen
  \bibfield  {author} {\bibinfo {author} {\bibfnamefont {A.~I.}\ \bibnamefont
  {Berdyugin}}, \bibinfo {author} {\bibfnamefont {B.}~\bibnamefont {Tsim}},
  \bibinfo {author} {\bibfnamefont {P.}~\bibnamefont {Kumaravadivel}}, \bibinfo
  {author} {\bibfnamefont {S.~G.}\ \bibnamefont {Xu}}, \bibinfo {author}
  {\bibfnamefont {A.}~\bibnamefont {Ceferino}}, \bibinfo {author}
  {\bibfnamefont {A.}~\bibnamefont {Knothe}}, \bibinfo {author} {\bibfnamefont
  {R.~K.}\ \bibnamefont {Kumar}}, \bibinfo {author} {\bibfnamefont
  {T.}~\bibnamefont {Taniguchi}}, \bibinfo {author} {\bibfnamefont
  {K.}~\bibnamefont {Watanabe}}, \bibinfo {author} {\bibfnamefont {A.~K.}\
  \bibnamefont {Geim}}, \bibinfo {author} {\bibfnamefont {I.~V.}\ \bibnamefont
  {Grigorieva}},\ and\ \bibinfo {author} {\bibfnamefont {V.~I.}\ \bibnamefont
  {Fal'ko}},\ }\bibfield  {title} {\bibinfo {title} {Minibands in twisted
  bilayer graphene probed by magnetic focusing},\ }\href
  {https://doi.org/10.1126/sciadv.aay7838} {\bibfield  {journal} {\bibinfo
  {journal} {Science Advances}\ }\textbf {\bibinfo {volume} {6}},\ \bibinfo
  {pages} {eaay7838} (\bibinfo {year} {2020})}\BibitemShut {NoStop}%
\bibitem [{\citenamefont {Eich}\ \emph {et~al.}(2018)\citenamefont {Eich},
  \citenamefont {Pisoni}, \citenamefont {Pally}, \citenamefont {Overweg},
  \citenamefont {Kurzmann}, \citenamefont {Lee}, \citenamefont {Rickhaus},
  \citenamefont {Watanabe}, \citenamefont {Taniguchi}, \citenamefont
  {Ensslin},\ and\ \citenamefont {Ihn}}]{eichCoupledQuantumDots2018}%
  \BibitemOpen
  \bibfield  {author} {\bibinfo {author} {\bibfnamefont {M.}~\bibnamefont
  {Eich}}, \bibinfo {author} {\bibfnamefont {R.}~\bibnamefont {Pisoni}},
  \bibinfo {author} {\bibfnamefont {A.}~\bibnamefont {Pally}}, \bibinfo
  {author} {\bibfnamefont {H.}~\bibnamefont {Overweg}}, \bibinfo {author}
  {\bibfnamefont {A.}~\bibnamefont {Kurzmann}}, \bibinfo {author}
  {\bibfnamefont {Y.}~\bibnamefont {Lee}}, \bibinfo {author} {\bibfnamefont
  {P.}~\bibnamefont {Rickhaus}}, \bibinfo {author} {\bibfnamefont
  {K.}~\bibnamefont {Watanabe}}, \bibinfo {author} {\bibfnamefont
  {T.}~\bibnamefont {Taniguchi}}, \bibinfo {author} {\bibfnamefont
  {K.}~\bibnamefont {Ensslin}},\ and\ \bibinfo {author} {\bibfnamefont
  {T.}~\bibnamefont {Ihn}},\ }\bibfield  {title} {\bibinfo {title} {Coupled
  {{Quantum Dots}} in {{Bilayer Graphene}}},\ }\href
  {https://doi.org/10.1021/acs.nanolett.8b01859} {\bibfield  {journal}
  {\bibinfo  {journal} {Nano Letters}\ }\textbf {\bibinfo {volume} {18}},\
  \bibinfo {pages} {5042} (\bibinfo {year} {2018})}\BibitemShut {NoStop}%
\bibitem [{\citenamefont {Overweg}\ \emph
  {et~al.}(2018{\natexlab{a}})\citenamefont {Overweg}, \citenamefont
  {Eggimann}, \citenamefont {Chen}, \citenamefont {Slizovskiy}, \citenamefont
  {Eich}, \citenamefont {Pisoni}, \citenamefont {Lee}, \citenamefont
  {Rickhaus}, \citenamefont {Watanabe}, \citenamefont {Taniguchi},
  \citenamefont {Fal'ko}, \citenamefont {Ihn},\ and\ \citenamefont
  {Ensslin}}]{overwegElectrostaticallyInducedQuantum2018}%
  \BibitemOpen
  \bibfield  {author} {\bibinfo {author} {\bibfnamefont {H.}~\bibnamefont
  {Overweg}}, \bibinfo {author} {\bibfnamefont {H.}~\bibnamefont {Eggimann}},
  \bibinfo {author} {\bibfnamefont {X.}~\bibnamefont {Chen}}, \bibinfo {author}
  {\bibfnamefont {S.}~\bibnamefont {Slizovskiy}}, \bibinfo {author}
  {\bibfnamefont {M.}~\bibnamefont {Eich}}, \bibinfo {author} {\bibfnamefont
  {R.}~\bibnamefont {Pisoni}}, \bibinfo {author} {\bibfnamefont
  {Y.}~\bibnamefont {Lee}}, \bibinfo {author} {\bibfnamefont {P.}~\bibnamefont
  {Rickhaus}}, \bibinfo {author} {\bibfnamefont {K.}~\bibnamefont {Watanabe}},
  \bibinfo {author} {\bibfnamefont {T.}~\bibnamefont {Taniguchi}}, \bibinfo
  {author} {\bibfnamefont {V.}~\bibnamefont {Fal'ko}}, \bibinfo {author}
  {\bibfnamefont {T.}~\bibnamefont {Ihn}},\ and\ \bibinfo {author}
  {\bibfnamefont {K.}~\bibnamefont {Ensslin}},\ }\bibfield  {title} {\bibinfo
  {title} {Electrostatically {{Induced Quantum Point Contacts}} in {{Bilayer
  Graphene}}},\ }\href {https://doi.org/10.1021/acs.nanolett.7b04666}
  {\bibfield  {journal} {\bibinfo  {journal} {Nano Letters}\ }\textbf {\bibinfo
  {volume} {18}},\ \bibinfo {pages} {553} (\bibinfo {year}
  {2018}{\natexlab{a}})}\BibitemShut {NoStop}%
\bibitem [{\citenamefont {Overweg}\ \emph
  {et~al.}(2018{\natexlab{b}})\citenamefont {Overweg}, \citenamefont {Knothe},
  \citenamefont {Fabian}, \citenamefont {Linhart}, \citenamefont {Rickhaus},
  \citenamefont {Wernli}, \citenamefont {Watanabe}, \citenamefont {Taniguchi},
  \citenamefont {S{\'a}nchez}, \citenamefont {Burgd{\"o}rfer}, \citenamefont
  {Libisch}, \citenamefont {Fal'ko}, \citenamefont {Ensslin},\ and\
  \citenamefont {Ihn}}]{overwegTopologicallyNontrivialValley2018}%
  \BibitemOpen
  \bibfield  {author} {\bibinfo {author} {\bibfnamefont {H.}~\bibnamefont
  {Overweg}}, \bibinfo {author} {\bibfnamefont {A.}~\bibnamefont {Knothe}},
  \bibinfo {author} {\bibfnamefont {T.}~\bibnamefont {Fabian}}, \bibinfo
  {author} {\bibfnamefont {L.}~\bibnamefont {Linhart}}, \bibinfo {author}
  {\bibfnamefont {P.}~\bibnamefont {Rickhaus}}, \bibinfo {author}
  {\bibfnamefont {L.}~\bibnamefont {Wernli}}, \bibinfo {author} {\bibfnamefont
  {K.}~\bibnamefont {Watanabe}}, \bibinfo {author} {\bibfnamefont
  {T.}~\bibnamefont {Taniguchi}}, \bibinfo {author} {\bibfnamefont
  {D.}~\bibnamefont {S{\'a}nchez}}, \bibinfo {author} {\bibfnamefont
  {J.}~\bibnamefont {Burgd{\"o}rfer}}, \bibinfo {author} {\bibfnamefont
  {F.}~\bibnamefont {Libisch}}, \bibinfo {author} {\bibfnamefont {V.~I.}\
  \bibnamefont {Fal'ko}}, \bibinfo {author} {\bibfnamefont {K.}~\bibnamefont
  {Ensslin}},\ and\ \bibinfo {author} {\bibfnamefont {T.}~\bibnamefont {Ihn}},\
  }\bibfield  {title} {\bibinfo {title} {Topologically {{Nontrivial Valley
  States}} in {{Bilayer Graphene Quantum Point Contacts}}},\ }\href
  {https://doi.org/10.1103/PhysRevLett.121.257702} {\bibfield  {journal}
  {\bibinfo  {journal} {Physical Review Letters}\ }\textbf {\bibinfo {volume}
  {121}},\ \bibinfo {pages} {257702} (\bibinfo {year}
  {2018}{\natexlab{b}})}\BibitemShut {NoStop}%
\bibitem [{\citenamefont {Banszerus}\ \emph {et~al.}(2019)\citenamefont
  {Banszerus}, \citenamefont {Frohn}, \citenamefont {Fabian}, \citenamefont
  {Somanchi}, \citenamefont {Epping}, \citenamefont {M{\"u}ller}, \citenamefont
  {Neumaier}, \citenamefont {Watanabe}, \citenamefont {Taniguchi},
  \citenamefont {Libisch}, \citenamefont {Beschoten}, \citenamefont {Hassler},\
  and\ \citenamefont {Stampfer}}]{banszerusSpinpolarizedCurrentsBilayer2019}%
  \BibitemOpen
  \bibfield  {author} {\bibinfo {author} {\bibfnamefont {L.}~\bibnamefont
  {Banszerus}}, \bibinfo {author} {\bibfnamefont {B.}~\bibnamefont {Frohn}},
  \bibinfo {author} {\bibfnamefont {T.}~\bibnamefont {Fabian}}, \bibinfo
  {author} {\bibfnamefont {S.}~\bibnamefont {Somanchi}}, \bibinfo {author}
  {\bibfnamefont {A.}~\bibnamefont {Epping}}, \bibinfo {author} {\bibfnamefont
  {M.}~\bibnamefont {M{\"u}ller}}, \bibinfo {author} {\bibfnamefont
  {D.}~\bibnamefont {Neumaier}}, \bibinfo {author} {\bibfnamefont
  {K.}~\bibnamefont {Watanabe}}, \bibinfo {author} {\bibfnamefont
  {T.}~\bibnamefont {Taniguchi}}, \bibinfo {author} {\bibfnamefont
  {F.}~\bibnamefont {Libisch}}, \bibinfo {author} {\bibfnamefont
  {B.}~\bibnamefont {Beschoten}}, \bibinfo {author} {\bibfnamefont
  {F.}~\bibnamefont {Hassler}},\ and\ \bibinfo {author} {\bibfnamefont
  {C.}~\bibnamefont {Stampfer}},\ }\bibfield  {title} {\bibinfo {title}
  {Spin-polarized currents in bilayer graphene quantum point contacts},\
  }\href@noop {} {\bibfield  {journal} {\bibinfo  {journal} {arXiv:1911.13176
  [cond-mat]}\ } (\bibinfo {year} {2019})},\ \Eprint
  {https://arxiv.org/abs/1911.13176} {arXiv:1911.13176 [cond-mat]} \BibitemShut
  {NoStop}%
\bibitem [{\citenamefont {Banszerus}\ \emph {et~al.}(2018)\citenamefont
  {Banszerus}, \citenamefont {Frohn}, \citenamefont {Epping}, \citenamefont
  {Neumaier}, \citenamefont {Watanabe}, \citenamefont {Taniguchi},\ and\
  \citenamefont {Stampfer}}]{banszerusGateDefinedElectronHole2018}%
  \BibitemOpen
  \bibfield  {author} {\bibinfo {author} {\bibfnamefont {L.}~\bibnamefont
  {Banszerus}}, \bibinfo {author} {\bibfnamefont {B.}~\bibnamefont {Frohn}},
  \bibinfo {author} {\bibfnamefont {A.}~\bibnamefont {Epping}}, \bibinfo
  {author} {\bibfnamefont {D.}~\bibnamefont {Neumaier}}, \bibinfo {author}
  {\bibfnamefont {K.}~\bibnamefont {Watanabe}}, \bibinfo {author}
  {\bibfnamefont {T.}~\bibnamefont {Taniguchi}},\ and\ \bibinfo {author}
  {\bibfnamefont {C.}~\bibnamefont {Stampfer}},\ }\bibfield  {title} {\bibinfo
  {title} {Gate-{{Defined Electron}}\textendash{{Hole Double Dots}} in
  {{Bilayer Graphene}}},\ }\href {https://doi.org/10.1021/acs.nanolett.8b01303}
  {\bibfield  {journal} {\bibinfo  {journal} {Nano Letters}\ }\textbf {\bibinfo
  {volume} {18}},\ \bibinfo {pages} {4785} (\bibinfo {year}
  {2018})}\BibitemShut {NoStop}%
\bibitem [{\citenamefont {Lee}\ \emph {et~al.}(2020)\citenamefont {Lee},
  \citenamefont {Knothe}, \citenamefont {Overweg}, \citenamefont {Eich},
  \citenamefont {Gold}, \citenamefont {Kurzmann}, \citenamefont {Klasovika},
  \citenamefont {Taniguchi}, \citenamefont {Wantanabe}, \citenamefont {Fal'ko},
  \citenamefont {Ihn}, \citenamefont {Ensslin},\ and\ \citenamefont
  {Rickhaus}}]{leeTunableValleySplitting2020}%
  \BibitemOpen
  \bibfield  {author} {\bibinfo {author} {\bibfnamefont {Y.}~\bibnamefont
  {Lee}}, \bibinfo {author} {\bibfnamefont {A.}~\bibnamefont {Knothe}},
  \bibinfo {author} {\bibfnamefont {H.}~\bibnamefont {Overweg}}, \bibinfo
  {author} {\bibfnamefont {M.}~\bibnamefont {Eich}}, \bibinfo {author}
  {\bibfnamefont {C.}~\bibnamefont {Gold}}, \bibinfo {author} {\bibfnamefont
  {A.}~\bibnamefont {Kurzmann}}, \bibinfo {author} {\bibfnamefont
  {V.}~\bibnamefont {Klasovika}}, \bibinfo {author} {\bibfnamefont
  {T.}~\bibnamefont {Taniguchi}}, \bibinfo {author} {\bibfnamefont
  {K.}~\bibnamefont {Wantanabe}}, \bibinfo {author} {\bibfnamefont
  {V.}~\bibnamefont {Fal'ko}}, \bibinfo {author} {\bibfnamefont
  {T.}~\bibnamefont {Ihn}}, \bibinfo {author} {\bibfnamefont {K.}~\bibnamefont
  {Ensslin}},\ and\ \bibinfo {author} {\bibfnamefont {P.}~\bibnamefont
  {Rickhaus}},\ }\bibfield  {title} {\bibinfo {title} {Tunable {{Valley
  Splitting}} due to {{Topological Orbital Magnetic Moment}} in {{Bilayer
  Graphene Quantum Point Contacts}}},\ }\href
  {https://doi.org/10.1103/PhysRevLett.124.126802} {\bibfield  {journal}
  {\bibinfo  {journal} {Physical Review Letters}\ }\textbf {\bibinfo {volume}
  {124}},\ \bibinfo {pages} {126802} (\bibinfo {year} {2020})}\BibitemShut
  {NoStop}%
\bibitem [{\citenamefont {M{\"o}ller}\ \emph {et~al.}(2021)\citenamefont
  {M{\"o}ller}, \citenamefont {Banszerus}, \citenamefont {Knothe},
  \citenamefont {Steiner}, \citenamefont {Icking}, \citenamefont {Trellenkamp},
  \citenamefont {Lentz}, \citenamefont {Watanabe}, \citenamefont {Taniguchi},
  \citenamefont {Glazman}, \citenamefont {Fal'ko}, \citenamefont {Volk},\ and\
  \citenamefont {Stampfer}}]{mollerProbingTwoElectronMultiplets2021}%
  \BibitemOpen
  \bibfield  {author} {\bibinfo {author} {\bibfnamefont {S.}~\bibnamefont
  {M{\"o}ller}}, \bibinfo {author} {\bibfnamefont {L.}~\bibnamefont
  {Banszerus}}, \bibinfo {author} {\bibfnamefont {A.}~\bibnamefont {Knothe}},
  \bibinfo {author} {\bibfnamefont {C.}~\bibnamefont {Steiner}}, \bibinfo
  {author} {\bibfnamefont {E.}~\bibnamefont {Icking}}, \bibinfo {author}
  {\bibfnamefont {S.}~\bibnamefont {Trellenkamp}}, \bibinfo {author}
  {\bibfnamefont {F.}~\bibnamefont {Lentz}}, \bibinfo {author} {\bibfnamefont
  {K.}~\bibnamefont {Watanabe}}, \bibinfo {author} {\bibfnamefont
  {T.}~\bibnamefont {Taniguchi}}, \bibinfo {author} {\bibfnamefont {L.~I.}\
  \bibnamefont {Glazman}}, \bibinfo {author} {\bibfnamefont {V.~I.}\
  \bibnamefont {Fal'ko}}, \bibinfo {author} {\bibfnamefont {C.}~\bibnamefont
  {Volk}},\ and\ \bibinfo {author} {\bibfnamefont {C.}~\bibnamefont
  {Stampfer}},\ }\bibfield  {title} {\bibinfo {title} {Probing {{Two-Electron
  Multiplets}} in {{Bilayer Graphene Quantum Dots}}},\ }\href
  {https://doi.org/10.1103/PhysRevLett.127.256802} {\bibfield  {journal}
  {\bibinfo  {journal} {Physical Review Letters}\ }\textbf {\bibinfo {volume}
  {127}},\ \bibinfo {pages} {256802} (\bibinfo {year} {2021})}\BibitemShut
  {NoStop}%
\bibitem [{\citenamefont {Tong}\ \emph {et~al.}(2021)\citenamefont {Tong},
  \citenamefont {Garreis}, \citenamefont {Knothe}, \citenamefont {Eich},
  \citenamefont {Sacchi}, \citenamefont {Watanabe}, \citenamefont {Taniguchi},
  \citenamefont {Fal'ko}, \citenamefont {Ihn}, \citenamefont {Ensslin},\ and\
  \citenamefont {Kurzmann}}]{tongTunableValleySplitting2021}%
  \BibitemOpen
  \bibfield  {author} {\bibinfo {author} {\bibfnamefont {C.}~\bibnamefont
  {Tong}}, \bibinfo {author} {\bibfnamefont {R.}~\bibnamefont {Garreis}},
  \bibinfo {author} {\bibfnamefont {A.}~\bibnamefont {Knothe}}, \bibinfo
  {author} {\bibfnamefont {M.}~\bibnamefont {Eich}}, \bibinfo {author}
  {\bibfnamefont {A.}~\bibnamefont {Sacchi}}, \bibinfo {author} {\bibfnamefont
  {K.}~\bibnamefont {Watanabe}}, \bibinfo {author} {\bibfnamefont
  {T.}~\bibnamefont {Taniguchi}}, \bibinfo {author} {\bibfnamefont
  {V.}~\bibnamefont {Fal'ko}}, \bibinfo {author} {\bibfnamefont
  {T.}~\bibnamefont {Ihn}}, \bibinfo {author} {\bibfnamefont {K.}~\bibnamefont
  {Ensslin}},\ and\ \bibinfo {author} {\bibfnamefont {A.}~\bibnamefont
  {Kurzmann}},\ }\bibfield  {title} {\bibinfo {title} {Tunable {{Valley
  Splitting}} and {{Bipolar Operation}} in {{Graphene Quantum Dots}}},\ }\href
  {https://doi.org/10.1021/acs.nanolett.0c04343} {\bibfield  {journal}
  {\bibinfo  {journal} {Nano Letters}\ }\textbf {\bibinfo {volume} {21}},\
  \bibinfo {pages} {1068} (\bibinfo {year} {2021})}\BibitemShut {NoStop}%
\bibitem [{\citenamefont {Knothe}\ and\ \citenamefont
  {Fal'ko}(2020)}]{knotheQuartetStatesTwoelectron2020}%
  \BibitemOpen
  \bibfield  {author} {\bibinfo {author} {\bibfnamefont {A.}~\bibnamefont
  {Knothe}}\ and\ \bibinfo {author} {\bibfnamefont {V.}~\bibnamefont
  {Fal'ko}},\ }\bibfield  {title} {\bibinfo {title} {Quartet states in
  two-electron quantum dots in bilayer graphene},\ }\href
  {https://doi.org/10.1103/PhysRevB.101.235423} {\bibfield  {journal} {\bibinfo
   {journal} {Physical Review B}\ }\textbf {\bibinfo {volume} {101}},\ \bibinfo
  {pages} {235423} (\bibinfo {year} {2020})}\BibitemShut {NoStop}%
\bibitem [{\citenamefont {Lane}\ \emph {et~al.}(2019)\citenamefont {Lane},
  \citenamefont {Knothe},\ and\ \citenamefont
  {Fal'ko}}]{laneSemimetallicFeaturesQuantum2019}%
  \BibitemOpen
  \bibfield  {author} {\bibinfo {author} {\bibfnamefont {T.~L.~M.}\
  \bibnamefont {Lane}}, \bibinfo {author} {\bibfnamefont {A.}~\bibnamefont
  {Knothe}},\ and\ \bibinfo {author} {\bibfnamefont {V.~I.}\ \bibnamefont
  {Fal'ko}},\ }\bibfield  {title} {\bibinfo {title} {Semimetallic features in
  quantum transport through a gate-defined point contact in bilayer graphene},\
  }\href {https://doi.org/10.1103/PhysRevB.100.115427} {\bibfield  {journal}
  {\bibinfo  {journal} {Physical Review B}\ }\textbf {\bibinfo {volume}
  {100}},\ \bibinfo {pages} {115427} (\bibinfo {year} {2019})}\BibitemShut
  {NoStop}%
\bibitem [{\citenamefont {Garreis}\ \emph {et~al.}(2021)\citenamefont
  {Garreis}, \citenamefont {Knothe}, \citenamefont {Tong}, \citenamefont
  {Eich}, \citenamefont {Gold}, \citenamefont {Watanabe}, \citenamefont
  {Taniguchi}, \citenamefont {Fal'ko}, \citenamefont {Ihn}, \citenamefont
  {Ensslin},\ and\ \citenamefont {Kurzmann}}]{garreisShellFillingTrigonal2021}%
  \BibitemOpen
  \bibfield  {author} {\bibinfo {author} {\bibfnamefont {R.}~\bibnamefont
  {Garreis}}, \bibinfo {author} {\bibfnamefont {A.}~\bibnamefont {Knothe}},
  \bibinfo {author} {\bibfnamefont {C.}~\bibnamefont {Tong}}, \bibinfo {author}
  {\bibfnamefont {M.}~\bibnamefont {Eich}}, \bibinfo {author} {\bibfnamefont
  {C.}~\bibnamefont {Gold}}, \bibinfo {author} {\bibfnamefont {K.}~\bibnamefont
  {Watanabe}}, \bibinfo {author} {\bibfnamefont {T.}~\bibnamefont {Taniguchi}},
  \bibinfo {author} {\bibfnamefont {V.}~\bibnamefont {Fal'ko}}, \bibinfo
  {author} {\bibfnamefont {T.}~\bibnamefont {Ihn}}, \bibinfo {author}
  {\bibfnamefont {K.}~\bibnamefont {Ensslin}},\ and\ \bibinfo {author}
  {\bibfnamefont {A.}~\bibnamefont {Kurzmann}},\ }\bibfield  {title} {\bibinfo
  {title} {Shell {{Filling}} and {{Trigonal Warping}} in {{Graphene Quantum
  Dots}}},\ }\href {https://doi.org/10.1103/PhysRevLett.126.147703} {\bibfield
  {journal} {\bibinfo  {journal} {Physical Review Letters}\ }\textbf {\bibinfo
  {volume} {126}},\ \bibinfo {pages} {147703} (\bibinfo {year}
  {2021})}\BibitemShut {NoStop}%
\bibitem [{\citenamefont {Knothe}\ \emph {et~al.}(2022)\citenamefont {Knothe},
  \citenamefont {Glazman},\ and\ \citenamefont
  {Fal'ko}}]{knotheTunnelingTheoryBilayer2022}%
  \BibitemOpen
  \bibfield  {author} {\bibinfo {author} {\bibfnamefont {A.}~\bibnamefont
  {Knothe}}, \bibinfo {author} {\bibfnamefont {L.~I.}\ \bibnamefont
  {Glazman}},\ and\ \bibinfo {author} {\bibfnamefont {V.~I.}\ \bibnamefont
  {Fal'ko}},\ }\bibfield  {title} {\bibinfo {title} {Tunneling theory for a
  bilayer graphene quantum dot's single- and two-electron states},\ }\href
  {https://doi.org/10.1088/1367-2630/ac5d00} {\bibfield  {journal} {\bibinfo
  {journal} {New Journal of Physics}\ }\textbf {\bibinfo {volume} {24}},\
  \bibinfo {pages} {043003} (\bibinfo {year} {2022})}\BibitemShut {NoStop}%
\bibitem [{\citenamefont {Iwakiri}\ \emph {et~al.}(2022)\citenamefont
  {Iwakiri}, \citenamefont {{de Vries}}, \citenamefont {Portol{\'e}s},
  \citenamefont {Zheng}, \citenamefont {Taniguchi}, \citenamefont {Watanabe},
  \citenamefont {Ihn},\ and\ \citenamefont
  {Ensslin}}]{iwakiriGateDefinedElectronInterferometer2022}%
  \BibitemOpen
  \bibfield  {author} {\bibinfo {author} {\bibfnamefont {S.}~\bibnamefont
  {Iwakiri}}, \bibinfo {author} {\bibfnamefont {F.~K.}\ \bibnamefont {{de
  Vries}}}, \bibinfo {author} {\bibfnamefont {E.}~\bibnamefont {Portol{\'e}s}},
  \bibinfo {author} {\bibfnamefont {G.}~\bibnamefont {Zheng}}, \bibinfo
  {author} {\bibfnamefont {T.}~\bibnamefont {Taniguchi}}, \bibinfo {author}
  {\bibfnamefont {K.}~\bibnamefont {Watanabe}}, \bibinfo {author}
  {\bibfnamefont {T.}~\bibnamefont {Ihn}},\ and\ \bibinfo {author}
  {\bibfnamefont {K.}~\bibnamefont {Ensslin}},\ }\bibfield  {title} {\bibinfo
  {title} {Gate-{{Defined Electron Interferometer}} in {{Bilayer Graphene}}},\
  }\href {https://doi.org/10.1021/acs.nanolett.2c01874} {\bibfield  {journal}
  {\bibinfo  {journal} {Nano Letters}\ }\textbf {\bibinfo {volume} {22}},\
  \bibinfo {pages} {6292} (\bibinfo {year} {2022})}\BibitemShut {NoStop}%
\bibitem [{\citenamefont {{Rodan-Legrain}}\ \emph {et~al.}(2021)\citenamefont
  {{Rodan-Legrain}}, \citenamefont {Cao}, \citenamefont {Park}, \citenamefont
  {{de la Barrera}}, \citenamefont {Randeria}, \citenamefont {Watanabe},
  \citenamefont {Taniguchi},\ and\ \citenamefont
  {{Jarillo-Herrero}}}]{rodan-legrainHighlyTunableJunctions2021}%
  \BibitemOpen
  \bibfield  {author} {\bibinfo {author} {\bibfnamefont {D.}~\bibnamefont
  {{Rodan-Legrain}}}, \bibinfo {author} {\bibfnamefont {Y.}~\bibnamefont
  {Cao}}, \bibinfo {author} {\bibfnamefont {J.~M.}\ \bibnamefont {Park}},
  \bibinfo {author} {\bibfnamefont {S.~C.}\ \bibnamefont {{de la Barrera}}},
  \bibinfo {author} {\bibfnamefont {M.~T.}\ \bibnamefont {Randeria}}, \bibinfo
  {author} {\bibfnamefont {K.}~\bibnamefont {Watanabe}}, \bibinfo {author}
  {\bibfnamefont {T.}~\bibnamefont {Taniguchi}},\ and\ \bibinfo {author}
  {\bibfnamefont {P.}~\bibnamefont {{Jarillo-Herrero}}},\ }\bibfield  {title}
  {\bibinfo {title} {Highly tunable junctions and non-local {{Josephson}}
  effect in magic-angle graphene tunnelling devices},\ }\href
  {https://doi.org/10.1038/s41565-021-00894-4} {\bibfield  {journal} {\bibinfo
  {journal} {Nature Nanotechnology}\ }\textbf {\bibinfo {volume} {16}},\
  \bibinfo {pages} {769} (\bibinfo {year} {2021})}\BibitemShut {NoStop}%
\bibitem [{\citenamefont {{de Vries}}\ \emph {et~al.}(2021)\citenamefont {{de
  Vries}}, \citenamefont {Portol{\'e}s}, \citenamefont {Zheng}, \citenamefont
  {Taniguchi}, \citenamefont {Watanabe}, \citenamefont {Ihn}, \citenamefont
  {Ensslin},\ and\ \citenamefont
  {Rickhaus}}]{devriesGatedefinedJosephsonJunctions2021}%
  \BibitemOpen
  \bibfield  {author} {\bibinfo {author} {\bibfnamefont {F.~K.}\ \bibnamefont
  {{de Vries}}}, \bibinfo {author} {\bibfnamefont {E.}~\bibnamefont
  {Portol{\'e}s}}, \bibinfo {author} {\bibfnamefont {G.}~\bibnamefont {Zheng}},
  \bibinfo {author} {\bibfnamefont {T.}~\bibnamefont {Taniguchi}}, \bibinfo
  {author} {\bibfnamefont {K.}~\bibnamefont {Watanabe}}, \bibinfo {author}
  {\bibfnamefont {T.}~\bibnamefont {Ihn}}, \bibinfo {author} {\bibfnamefont
  {K.}~\bibnamefont {Ensslin}},\ and\ \bibinfo {author} {\bibfnamefont
  {P.}~\bibnamefont {Rickhaus}},\ }\bibfield  {title} {\bibinfo {title}
  {Gate-defined {{Josephson}} junctions in magic-angle twisted bilayer
  graphene},\ }\href {https://doi.org/10.1038/s41565-021-00896-2} {\bibfield
  {journal} {\bibinfo  {journal} {Nature Nanotechnology}\ }\textbf {\bibinfo
  {volume} {16}},\ \bibinfo {pages} {760} (\bibinfo {year} {2021})}\BibitemShut
  {NoStop}%
\bibitem [{\citenamefont {Portol{\'e}s}\ \emph {et~al.}(2022)\citenamefont
  {Portol{\'e}s}, \citenamefont {Iwakiri}, \citenamefont {Zheng}, \citenamefont
  {Rickhaus}, \citenamefont {Taniguchi}, \citenamefont {Watanabe},
  \citenamefont {Ihn}, \citenamefont {Ensslin},\ and\ \citenamefont {{de
  Vries}}}]{portolesTunableMonolithicSQUID2022}%
  \BibitemOpen
  \bibfield  {author} {\bibinfo {author} {\bibfnamefont {E.}~\bibnamefont
  {Portol{\'e}s}}, \bibinfo {author} {\bibfnamefont {S.}~\bibnamefont
  {Iwakiri}}, \bibinfo {author} {\bibfnamefont {G.}~\bibnamefont {Zheng}},
  \bibinfo {author} {\bibfnamefont {P.}~\bibnamefont {Rickhaus}}, \bibinfo
  {author} {\bibfnamefont {T.}~\bibnamefont {Taniguchi}}, \bibinfo {author}
  {\bibfnamefont {K.}~\bibnamefont {Watanabe}}, \bibinfo {author}
  {\bibfnamefont {T.}~\bibnamefont {Ihn}}, \bibinfo {author} {\bibfnamefont
  {K.}~\bibnamefont {Ensslin}},\ and\ \bibinfo {author} {\bibfnamefont {F.~K.}\
  \bibnamefont {{de Vries}}},\ }\href
  {https://doi.org/10.48550/arXiv.2201.13276} {\bibinfo {title} {A {{Tunable
  Monolithic SQUID}} in {{Twisted Bilayer Graphene}}}} (\bibinfo {year}
  {2022}),\ \Eprint {https://arxiv.org/abs/2201.13276} {arXiv:2201.13276
  [cond-mat]} \BibitemShut {NoStop}%
\bibitem [{\citenamefont {Dauber}\ \emph {et~al.}(2021)\citenamefont {Dauber},
  \citenamefont {Reijnders}, \citenamefont {Banszerus}, \citenamefont {Epping},
  \citenamefont {Watanabe}, \citenamefont {Taniguchi}, \citenamefont
  {Katsnelson}, \citenamefont {Hassler},\ and\ \citenamefont
  {Stampfer}}]{dauberExploitingAharonovBohmOscillations2021}%
  \BibitemOpen
  \bibfield  {author} {\bibinfo {author} {\bibfnamefont {J.}~\bibnamefont
  {Dauber}}, \bibinfo {author} {\bibfnamefont {K.~J.~A.}\ \bibnamefont
  {Reijnders}}, \bibinfo {author} {\bibfnamefont {L.}~\bibnamefont
  {Banszerus}}, \bibinfo {author} {\bibfnamefont {A.}~\bibnamefont {Epping}},
  \bibinfo {author} {\bibfnamefont {K.}~\bibnamefont {Watanabe}}, \bibinfo
  {author} {\bibfnamefont {T.}~\bibnamefont {Taniguchi}}, \bibinfo {author}
  {\bibfnamefont {M.~I.}\ \bibnamefont {Katsnelson}}, \bibinfo {author}
  {\bibfnamefont {F.}~\bibnamefont {Hassler}},\ and\ \bibinfo {author}
  {\bibfnamefont {C.}~\bibnamefont {Stampfer}},\ }\href
  {https://doi.org/10.48550/arXiv.2008.02556} {\bibinfo {title} {Exploiting
  {{Aharonov-Bohm}} oscillations to probe {{Klein}} tunneling in tunable
  pn-junctions in graphene}} (\bibinfo {year} {2021}),\ \Eprint
  {https://arxiv.org/abs/2008.02556} {arXiv:2008.02556 [cond-mat]} \BibitemShut
  {NoStop}%
\bibitem [{\citenamefont {Elahi}\ \emph {et~al.}(2022)\citenamefont {Elahi},
  \citenamefont {Zeng}, \citenamefont {Dean},\ and\ \citenamefont
  {Ghosh}}]{elahiDirectEvidenceKleinantiKlein2022}%
  \BibitemOpen
  \bibfield  {author} {\bibinfo {author} {\bibfnamefont {M.~M.}\ \bibnamefont
  {Elahi}}, \bibinfo {author} {\bibfnamefont {Y.}~\bibnamefont {Zeng}},
  \bibinfo {author} {\bibfnamefont {C.~R.}\ \bibnamefont {Dean}},\ and\
  \bibinfo {author} {\bibfnamefont {A.~W.}\ \bibnamefont {Ghosh}},\ }\href
  {https://doi.org/10.48550/arXiv.2210.10429} {\bibinfo {title} {Direct
  evidence of {{Klein-antiKlein}} tunneling of graphitic electrons in a
  {{Corbino}} geometry}} (\bibinfo {year} {2022}),\ \Eprint
  {https://arxiv.org/abs/2210.10429} {arXiv:2210.10429 [cond-mat]} \BibitemShut
  {NoStop}%
\bibitem [{\citenamefont {Cheianov}\ \emph {et~al.}(2007)\citenamefont
  {Cheianov}, \citenamefont {Fal'ko},\ and\ \citenamefont
  {Altshuler}}]{cheianovFocusingElectronFlow2007}%
  \BibitemOpen
  \bibfield  {author} {\bibinfo {author} {\bibfnamefont {V.~V.}\ \bibnamefont
  {Cheianov}}, \bibinfo {author} {\bibfnamefont {V.}~\bibnamefont {Fal'ko}},\
  and\ \bibinfo {author} {\bibfnamefont {B.~L.}\ \bibnamefont {Altshuler}},\
  }\bibfield  {title} {\bibinfo {title} {The {{Focusing}} of {{Electron Flow}}
  and a {{Veselago Lens}} in {{Graphene}} p-n {{Junctions}}},\ }\bibfield
  {journal} {\bibinfo  {journal} {Science}\ }\href
  {https://doi.org/10.1126/science.1138020} {10.1126/science.1138020} (\bibinfo
  {year} {2007})\BibitemShut {NoStop}%
\bibitem [{\citenamefont {P{\'e}terfalvi}\ \emph {et~al.}(2009)\citenamefont
  {P{\'e}terfalvi}, \citenamefont {P{\'a}lyi},\ and\ \citenamefont
  {Cserti}}]{peterfalviElectronFlowCircular2009}%
  \BibitemOpen
  \bibfield  {author} {\bibinfo {author} {\bibfnamefont {C.}~\bibnamefont
  {P{\'e}terfalvi}}, \bibinfo {author} {\bibfnamefont {A.}~\bibnamefont
  {P{\'a}lyi}},\ and\ \bibinfo {author} {\bibfnamefont {J.}~\bibnamefont
  {Cserti}},\ }\bibfield  {title} {\bibinfo {title} {Electron flow in circular
  \$n\textbackslash text\{\textbackslash ensuremath\{-\}\}p\$ junctions of
  bilayer graphene},\ }\href {https://doi.org/10.1103/PhysRevB.80.075416}
  {\bibfield  {journal} {\bibinfo  {journal} {Physical Review B}\ }\textbf
  {\bibinfo {volume} {80}},\ \bibinfo {pages} {075416} (\bibinfo {year}
  {2009})}\BibitemShut {NoStop}%
\bibitem [{\citenamefont {Xu}\ \emph {et~al.}(2021)\citenamefont {Xu},
  \citenamefont {Huang},\ and\ \citenamefont
  {Lai}}]{xuRelativisticQuantumChaos2021}%
  \BibitemOpen
  \bibfield  {author} {\bibinfo {author} {\bibfnamefont {H.-Y.}\ \bibnamefont
  {Xu}}, \bibinfo {author} {\bibfnamefont {L.}~\bibnamefont {Huang}},\ and\
  \bibinfo {author} {\bibfnamefont {Y.-C.}\ \bibnamefont {Lai}},\ }\bibfield
  {title} {\bibinfo {title} {Relativistic quantum chaos in graphene},\ }\href
  {https://doi.org/10.1063/PT.3.4679} {\bibfield  {journal} {\bibinfo
  {journal} {Physics Today}\ }\textbf {\bibinfo {volume} {74}},\ \bibinfo
  {pages} {44} (\bibinfo {year} {2021})}\BibitemShut {NoStop}%
\bibitem [{\citenamefont {Xu}\ \emph {et~al.}(2018)\citenamefont {Xu},
  \citenamefont {Wang}, \citenamefont {Huang},\ and\ \citenamefont
  {Lai}}]{xuChaosDiracElectron2018}%
  \BibitemOpen
  \bibfield  {author} {\bibinfo {author} {\bibfnamefont {H.-Y.}\ \bibnamefont
  {Xu}}, \bibinfo {author} {\bibfnamefont {G.-L.}\ \bibnamefont {Wang}},
  \bibinfo {author} {\bibfnamefont {L.}~\bibnamefont {Huang}},\ and\ \bibinfo
  {author} {\bibfnamefont {Y.-C.}\ \bibnamefont {Lai}},\ }\bibfield  {title}
  {\bibinfo {title} {Chaos in {{Dirac Electron Optics}}: {{Emergence}} of a
  {{Relativistic Quantum Chimera}}},\ }\href
  {https://doi.org/10.1103/PhysRevLett.120.124101} {\bibfield  {journal}
  {\bibinfo  {journal} {Physical Review Letters}\ }\textbf {\bibinfo {volume}
  {120}},\ \bibinfo {pages} {124101} (\bibinfo {year} {2018})}\BibitemShut
  {NoStop}%
\bibitem [{\citenamefont {Schneider}\ and\ \citenamefont
  {Brouwer}(2011)}]{schneiderResonantScatteringGraphene2011}%
  \BibitemOpen
  \bibfield  {author} {\bibinfo {author} {\bibfnamefont {M.}~\bibnamefont
  {Schneider}}\ and\ \bibinfo {author} {\bibfnamefont {P.~W.}\ \bibnamefont
  {Brouwer}},\ }\bibfield  {title} {\bibinfo {title} {Resonant scattering in
  graphene with a gate-defined chaotic quantum dot},\ }\href
  {https://doi.org/10.1103/PhysRevB.84.115440} {\bibfield  {journal} {\bibinfo
  {journal} {Physical Review B}\ }\textbf {\bibinfo {volume} {84}},\ \bibinfo
  {pages} {115440} (\bibinfo {year} {2011})}\BibitemShut {NoStop}%
\bibitem [{\citenamefont {Schneider}\ and\ \citenamefont
  {Brouwer}(2014)}]{schneiderDensityStatesProbe2014}%
  \BibitemOpen
  \bibfield  {author} {\bibinfo {author} {\bibfnamefont {M.}~\bibnamefont
  {Schneider}}\ and\ \bibinfo {author} {\bibfnamefont {P.~W.}\ \bibnamefont
  {Brouwer}},\ }\bibfield  {title} {\bibinfo {title} {Density of states as a
  probe of electrostatic confinement in graphene},\ }\href
  {https://doi.org/10.1103/PhysRevB.89.205437} {\bibfield  {journal} {\bibinfo
  {journal} {Physical Review B}\ }\textbf {\bibinfo {volume} {89}},\ \bibinfo
  {pages} {205437} (\bibinfo {year} {2014})}\BibitemShut {NoStop}%
\bibitem [{\citenamefont {Ponomarenko}\ \emph {et~al.}(2008)\citenamefont
  {Ponomarenko}, \citenamefont {Schedin}, \citenamefont {Katsnelson},
  \citenamefont {Yang}, \citenamefont {Hill}, \citenamefont {Novoselov},\ and\
  \citenamefont {Geim}}]{ponomarenkoChaoticDiracBilliard2008}%
  \BibitemOpen
  \bibfield  {author} {\bibinfo {author} {\bibfnamefont {L.~A.}\ \bibnamefont
  {Ponomarenko}}, \bibinfo {author} {\bibfnamefont {F.}~\bibnamefont
  {Schedin}}, \bibinfo {author} {\bibfnamefont {M.~I.}\ \bibnamefont
  {Katsnelson}}, \bibinfo {author} {\bibfnamefont {R.}~\bibnamefont {Yang}},
  \bibinfo {author} {\bibfnamefont {E.~W.}\ \bibnamefont {Hill}}, \bibinfo
  {author} {\bibfnamefont {K.~S.}\ \bibnamefont {Novoselov}},\ and\ \bibinfo
  {author} {\bibfnamefont {A.~K.}\ \bibnamefont {Geim}},\ }\bibfield  {title}
  {\bibinfo {title} {Chaotic {{Dirac Billiard}} in {{Graphene Quantum Dots}}},\
  }\href {https://doi.org/10.1126/science.1154663} {\bibfield  {journal}
  {\bibinfo  {journal} {Science}\ }\textbf {\bibinfo {volume} {320}},\ \bibinfo
  {pages} {356} (\bibinfo {year} {2008})}\BibitemShut {NoStop}%
\bibitem [{\citenamefont {Huang}\ and\ \citenamefont
  {Lai}(2020)}]{huangPerspectivesRelativisticQuantum2020}%
  \BibitemOpen
  \bibfield  {author} {\bibinfo {author} {\bibfnamefont {L.}~\bibnamefont
  {Huang}}\ and\ \bibinfo {author} {\bibfnamefont {Y.-C.}\ \bibnamefont
  {Lai}},\ }\bibfield  {title} {\bibinfo {title} {Perspectives on relativistic
  quantum chaos},\ }\href {https://doi.org/10.1088/1572-9494/ab6909} {\bibfield
   {journal} {\bibinfo  {journal} {Communications in Theoretical Physics}\
  }\textbf {\bibinfo {volume} {72}},\ \bibinfo {pages} {047601} (\bibinfo
  {year} {2020})}\BibitemShut {NoStop}%
\bibitem [{\citenamefont {Heinl}\ \emph {et~al.}(2013)\citenamefont {Heinl},
  \citenamefont {Schneider},\ and\ \citenamefont
  {Brouwer}}]{heinlInterplayAharonovBohmBerry2013}%
  \BibitemOpen
  \bibfield  {author} {\bibinfo {author} {\bibfnamefont {J.}~\bibnamefont
  {Heinl}}, \bibinfo {author} {\bibfnamefont {M.}~\bibnamefont {Schneider}},\
  and\ \bibinfo {author} {\bibfnamefont {P.~W.}\ \bibnamefont {Brouwer}},\
  }\bibfield  {title} {\bibinfo {title} {Interplay of {{Aharonov-Bohm}} and
  {{Berry}} phases in gate-defined graphene quantum dots},\ }\href
  {https://doi.org/10.1103/PhysRevB.87.245426} {\bibfield  {journal} {\bibinfo
  {journal} {Physical Review B}\ }\textbf {\bibinfo {volume} {87}},\ \bibinfo
  {pages} {245426} (\bibinfo {year} {2013})}\BibitemShut {NoStop}%
\bibitem [{\citenamefont {Han}\ \emph {et~al.}(2018)\citenamefont {Han},
  \citenamefont {Wang}, \citenamefont {Xu}, \citenamefont {Huang},\ and\
  \citenamefont {Lai}}]{hanDecaySemiclassicalMassless2018}%
  \BibitemOpen
  \bibfield  {author} {\bibinfo {author} {\bibfnamefont {C.-D.}\ \bibnamefont
  {Han}}, \bibinfo {author} {\bibfnamefont {C.-Z.}\ \bibnamefont {Wang}},
  \bibinfo {author} {\bibfnamefont {H.-Y.}\ \bibnamefont {Xu}}, \bibinfo
  {author} {\bibfnamefont {D.}~\bibnamefont {Huang}},\ and\ \bibinfo {author}
  {\bibfnamefont {Y.-C.}\ \bibnamefont {Lai}},\ }\bibfield  {title} {\bibinfo
  {title} {Decay of semiclassical massless {{Dirac}} fermions from integrable
  and chaotic cavities},\ }\href {https://doi.org/10.1103/PhysRevB.98.104308}
  {\bibfield  {journal} {\bibinfo  {journal} {Physical Review B}\ }\textbf
  {\bibinfo {volume} {98}},\ \bibinfo {pages} {104308} (\bibinfo {year}
  {2018})}\BibitemShut {NoStop}%
\bibitem [{\citenamefont {Ge}\ \emph {et~al.}(2021)\citenamefont {Ge},
  \citenamefont {Wong}, \citenamefont {Lee}, \citenamefont {Joucken},
  \citenamefont {{Quezada-Lopez}}, \citenamefont {Kahn}, \citenamefont {Tsai},
  \citenamefont {Taniguchi}, \citenamefont {Watanabe}, \citenamefont {Wang},
  \citenamefont {Zettl}, \citenamefont {Crommie},\ and\ \citenamefont
  {Velasco}}]{geImagingQuantumInterference2021}%
  \BibitemOpen
  \bibfield  {author} {\bibinfo {author} {\bibfnamefont {Z.}~\bibnamefont
  {Ge}}, \bibinfo {author} {\bibfnamefont {D.}~\bibnamefont {Wong}}, \bibinfo
  {author} {\bibfnamefont {J.}~\bibnamefont {Lee}}, \bibinfo {author}
  {\bibfnamefont {F.}~\bibnamefont {Joucken}}, \bibinfo {author} {\bibfnamefont
  {E.~A.}\ \bibnamefont {{Quezada-Lopez}}}, \bibinfo {author} {\bibfnamefont
  {S.}~\bibnamefont {Kahn}}, \bibinfo {author} {\bibfnamefont {H.-Z.}\
  \bibnamefont {Tsai}}, \bibinfo {author} {\bibfnamefont {T.}~\bibnamefont
  {Taniguchi}}, \bibinfo {author} {\bibfnamefont {K.}~\bibnamefont {Watanabe}},
  \bibinfo {author} {\bibfnamefont {F.}~\bibnamefont {Wang}}, \bibinfo {author}
  {\bibfnamefont {A.}~\bibnamefont {Zettl}}, \bibinfo {author} {\bibfnamefont
  {M.~F.}\ \bibnamefont {Crommie}},\ and\ \bibinfo {author} {\bibfnamefont
  {J.~J.}\ \bibnamefont {Velasco}},\ }\bibfield  {title} {\bibinfo {title}
  {Imaging {{Quantum Interference}} in {{Stadium-Shaped Monolayer}} and
  {{Bilayer Graphene Quantum Dots}}},\ }\href
  {https://doi.org/10.1021/acs.nanolett.1c02271} {\bibfield  {journal}
  {\bibinfo  {journal} {Nano Letters}\ }\textbf {\bibinfo {volume} {21}},\
  \bibinfo {pages} {8993} (\bibinfo {year} {2021})}\BibitemShut {NoStop}%
\bibitem [{\citenamefont {Brun}\ \emph {et~al.}(2022)\citenamefont {Brun},
  \citenamefont {Nguyen}, \citenamefont {Moreau}, \citenamefont {Somanchi},
  \citenamefont {Watanabe}, \citenamefont {Taniguchi}, \citenamefont
  {Charlier}, \citenamefont {Stampfer},\ and\ \citenamefont
  {Hackens}}]{brunGrapheneWhisperitronicsTransducing2022}%
  \BibitemOpen
  \bibfield  {author} {\bibinfo {author} {\bibfnamefont {B.}~\bibnamefont
  {Brun}}, \bibinfo {author} {\bibfnamefont {V.-H.}\ \bibnamefont {Nguyen}},
  \bibinfo {author} {\bibfnamefont {N.}~\bibnamefont {Moreau}}, \bibinfo
  {author} {\bibfnamefont {S.}~\bibnamefont {Somanchi}}, \bibinfo {author}
  {\bibfnamefont {K.}~\bibnamefont {Watanabe}}, \bibinfo {author}
  {\bibfnamefont {T.}~\bibnamefont {Taniguchi}}, \bibinfo {author}
  {\bibfnamefont {J.-C.}\ \bibnamefont {Charlier}}, \bibinfo {author}
  {\bibfnamefont {C.}~\bibnamefont {Stampfer}},\ and\ \bibinfo {author}
  {\bibfnamefont {B.}~\bibnamefont {Hackens}},\ }\bibfield  {title} {\bibinfo
  {title} {Graphene {{Whisperitronics}}: {{Transducing Whispering Gallery
  Modes}} into {{Electronic Transport}}},\ }\href
  {https://doi.org/10.1021/acs.nanolett.1c03451} {\bibfield  {journal}
  {\bibinfo  {journal} {Nano Letters}\ }\textbf {\bibinfo {volume} {22}},\
  \bibinfo {pages} {128} (\bibinfo {year} {2022})}\BibitemShut {NoStop}%
\bibitem [{\citenamefont {Miao}\ \emph {et~al.}(2007)\citenamefont {Miao},
  \citenamefont {Wijeratne}, \citenamefont {Zhang}, \citenamefont {Coskun},
  \citenamefont {Bao},\ and\ \citenamefont
  {Lau}}]{miaoPhaseCoherentTransportGraphene2007}%
  \BibitemOpen
  \bibfield  {author} {\bibinfo {author} {\bibfnamefont {F.}~\bibnamefont
  {Miao}}, \bibinfo {author} {\bibfnamefont {S.}~\bibnamefont {Wijeratne}},
  \bibinfo {author} {\bibfnamefont {Y.}~\bibnamefont {Zhang}}, \bibinfo
  {author} {\bibfnamefont {U.~C.}\ \bibnamefont {Coskun}}, \bibinfo {author}
  {\bibfnamefont {W.}~\bibnamefont {Bao}},\ and\ \bibinfo {author}
  {\bibfnamefont {C.~N.}\ \bibnamefont {Lau}},\ }\bibfield  {title} {\bibinfo
  {title} {Phase-{{Coherent Transport}} in {{Graphene Quantum Billiards}}},\
  }\href {https://doi.org/10.1126/science.1144359} {\bibfield  {journal}
  {\bibinfo  {journal} {Science}\ }\textbf {\bibinfo {volume} {317}},\ \bibinfo
  {pages} {1530} (\bibinfo {year} {2007})}\BibitemShut {NoStop}%
\bibitem [{\citenamefont {Bardarson}\ \emph {et~al.}(2009)\citenamefont
  {Bardarson}, \citenamefont {Titov},\ and\ \citenamefont
  {Brouwer}}]{bardarsonElectrostaticConfinementElectrons2009}%
  \BibitemOpen
  \bibfield  {author} {\bibinfo {author} {\bibfnamefont {J.~H.}\ \bibnamefont
  {Bardarson}}, \bibinfo {author} {\bibfnamefont {M.}~\bibnamefont {Titov}},\
  and\ \bibinfo {author} {\bibfnamefont {P.~W.}\ \bibnamefont {Brouwer}},\
  }\bibfield  {title} {\bibinfo {title} {Electrostatic {{Confinement}} of
  {{Electrons}} in an {{Integrable Graphene Quantum Dot}}},\ }\href
  {https://doi.org/10.1103/PhysRevLett.102.226803} {\bibfield  {journal}
  {\bibinfo  {journal} {Physical Review Letters}\ }\textbf {\bibinfo {volume}
  {102}},\ \bibinfo {pages} {226803} (\bibinfo {year} {2009})}\BibitemShut
  {NoStop}%
\bibitem [{\citenamefont {Schrepfer}\ \emph {et~al.}(2021)\citenamefont
  {Schrepfer}, \citenamefont {Chen}, \citenamefont {Liu}, \citenamefont
  {Richter},\ and\ \citenamefont
  {Hentschel}}]{schrepferDiracFermionOptics2021}%
  \BibitemOpen
  \bibfield  {author} {\bibinfo {author} {\bibfnamefont {J.-K.}\ \bibnamefont
  {Schrepfer}}, \bibinfo {author} {\bibfnamefont {S.-C.}\ \bibnamefont {Chen}},
  \bibinfo {author} {\bibfnamefont {M.-H.}\ \bibnamefont {Liu}}, \bibinfo
  {author} {\bibfnamefont {K.}~\bibnamefont {Richter}},\ and\ \bibinfo {author}
  {\bibfnamefont {M.}~\bibnamefont {Hentschel}},\ }\bibfield  {title} {\bibinfo
  {title} {Dirac fermion optics and directed emission from single- and bilayer
  graphene cavities},\ }\href {https://doi.org/10.1103/PhysRevB.104.155436}
  {\bibfield  {journal} {\bibinfo  {journal} {Physical Review B}\ }\textbf
  {\bibinfo {volume} {104}},\ \bibinfo {pages} {155436} (\bibinfo {year}
  {2021})}\BibitemShut {NoStop}%
\bibitem [{\citenamefont {Barbosa}\ \emph {et~al.}(2021)\citenamefont
  {Barbosa}, \citenamefont {Ramos},\ and\ \citenamefont
  {Ferreira}}]{PhysRevB.103.L081111}%
  \BibitemOpen
  \bibfield  {author} {\bibinfo {author} {\bibfnamefont {A.~L.~R.}\
  \bibnamefont {Barbosa}}, \bibinfo {author} {\bibfnamefont {J.~G. G.~S.}\
  \bibnamefont {Ramos}},\ and\ \bibinfo {author} {\bibfnamefont
  {A.}~\bibnamefont {Ferreira}},\ }\bibfield  {title} {\bibinfo {title} {Effect
  of proximity-induced spin-orbit coupling in graphene mesoscopic billiards},\
  }\href {https://doi.org/10.1103/PhysRevB.103.L081111} {\bibfield  {journal}
  {\bibinfo  {journal} {Phys. Rev. B}\ }\textbf {\bibinfo {volume} {103}},\
  \bibinfo {pages} {L081111} (\bibinfo {year} {2021})}\BibitemShut {NoStop}%
\bibitem [{\citenamefont {Katsnelson}\ \emph {et~al.}(2006)\citenamefont
  {Katsnelson}, \citenamefont {Novoselov},\ and\ \citenamefont
  {Geim}}]{katsnelsonChiralTunnellingKlein2006}%
  \BibitemOpen
  \bibfield  {author} {\bibinfo {author} {\bibfnamefont {M.~I.}\ \bibnamefont
  {Katsnelson}}, \bibinfo {author} {\bibfnamefont {K.~S.}\ \bibnamefont
  {Novoselov}},\ and\ \bibinfo {author} {\bibfnamefont {A.~K.}\ \bibnamefont
  {Geim}},\ }\bibfield  {title} {\bibinfo {title} {Chiral tunnelling and the
  {{Klein}} paradox in graphene},\ }\href {https://doi.org/10.1038/nphys384}
  {\bibfield  {journal} {\bibinfo  {journal} {Nature Physics}\ }\textbf
  {\bibinfo {volume} {2}},\ \bibinfo {pages} {620} (\bibinfo {year}
  {2006})}\BibitemShut {NoStop}%
\bibitem [{\citenamefont {Tudorovskiy}\ \emph {et~al.}(2012)\citenamefont
  {Tudorovskiy}, \citenamefont {Reijnders},\ and\ \citenamefont
  {Katsnelson}}]{tudorovskiyChiralTunnelingSinglelayer2012}%
  \BibitemOpen
  \bibfield  {author} {\bibinfo {author} {\bibfnamefont {T.}~\bibnamefont
  {Tudorovskiy}}, \bibinfo {author} {\bibfnamefont {K.~J.~A.}\ \bibnamefont
  {Reijnders}},\ and\ \bibinfo {author} {\bibfnamefont {M.~I.}\ \bibnamefont
  {Katsnelson}},\ }\bibfield  {title} {\bibinfo {title} {Chiral tunneling in
  single-layer and bilayer graphene},\ }\href
  {https://doi.org/10.1088/0031-8949/2012/T146/014010} {\bibfield  {journal}
  {\bibinfo  {journal} {Physica Scripta}\ }\textbf {\bibinfo {volume} {2012}},\
  \bibinfo {pages} {014010} (\bibinfo {year} {2012})}\BibitemShut {NoStop}%
\bibitem [{\citenamefont {Snyman}\ and\ \citenamefont
  {Beenakker}(2007)}]{snymanBallisticTransmissionGraphene2007}%
  \BibitemOpen
  \bibfield  {author} {\bibinfo {author} {\bibfnamefont {I.}~\bibnamefont
  {Snyman}}\ and\ \bibinfo {author} {\bibfnamefont {C.~W.~J.}\ \bibnamefont
  {Beenakker}},\ }\bibfield  {title} {\bibinfo {title} {Ballistic transmission
  through a graphene bilayer},\ }\href
  {https://doi.org/10.1103/PhysRevB.75.045322} {\bibfield  {journal} {\bibinfo
  {journal} {Physical Review B}\ }\textbf {\bibinfo {volume} {75}},\ \bibinfo
  {pages} {045322} (\bibinfo {year} {2007})}\BibitemShut {NoStop}%
\bibitem [{\citenamefont {Du}\ \emph {et~al.}(2018)\citenamefont {Du},
  \citenamefont {Liu}, \citenamefont {Mohrmann}, \citenamefont {Wu},
  \citenamefont {Krupke}, \citenamefont {{von L{\"o}hneysen}}, \citenamefont
  {Richter},\ and\ \citenamefont {Danneau}}]{duTuningAntiKleinKlein2018b}%
  \BibitemOpen
  \bibfield  {author} {\bibinfo {author} {\bibfnamefont {R.}~\bibnamefont
  {Du}}, \bibinfo {author} {\bibfnamefont {M.-H.}\ \bibnamefont {Liu}},
  \bibinfo {author} {\bibfnamefont {J.}~\bibnamefont {Mohrmann}}, \bibinfo
  {author} {\bibfnamefont {F.}~\bibnamefont {Wu}}, \bibinfo {author}
  {\bibfnamefont {R.}~\bibnamefont {Krupke}}, \bibinfo {author} {\bibfnamefont
  {H.}~\bibnamefont {{von L{\"o}hneysen}}}, \bibinfo {author} {\bibfnamefont
  {K.}~\bibnamefont {Richter}},\ and\ \bibinfo {author} {\bibfnamefont
  {R.}~\bibnamefont {Danneau}},\ }\bibfield  {title} {\bibinfo {title} {Tuning
  {{Anti-Klein}} to {{Klein Tunneling}} in {{Bilayer Graphene}}},\ }\href
  {https://doi.org/10.1103/PhysRevLett.121.127706} {\bibfield  {journal}
  {\bibinfo  {journal} {Physical Review Letters}\ }\textbf {\bibinfo {volume}
  {121}},\ \bibinfo {pages} {127706} (\bibinfo {year} {2018})}\BibitemShut
  {NoStop}%
\bibitem [{\citenamefont {Gradinar}\ \emph {et~al.}(2012)\citenamefont
  {Gradinar}, \citenamefont {Schomerus},\ and\ \citenamefont
  {Fal'ko}}]{gradinarConductanceAnomalyLifshitz2012}%
  \BibitemOpen
  \bibfield  {author} {\bibinfo {author} {\bibfnamefont {D.~A.}\ \bibnamefont
  {Gradinar}}, \bibinfo {author} {\bibfnamefont {H.}~\bibnamefont
  {Schomerus}},\ and\ \bibinfo {author} {\bibfnamefont {V.~I.}\ \bibnamefont
  {Fal'ko}},\ }\bibfield  {title} {\bibinfo {title} {Conductance anomaly near
  the {{Lifshitz}} transition in strained bilayer graphene},\ }\href
  {https://doi.org/10.1103/PhysRevB.85.165429} {\bibfield  {journal} {\bibinfo
  {journal} {Physical Review B}\ }\textbf {\bibinfo {volume} {85}},\ \bibinfo
  {pages} {165429} (\bibinfo {year} {2012})}\BibitemShut {NoStop}%
\bibitem [{\citenamefont {Nilsson}\ \emph {et~al.}(2007)\citenamefont
  {Nilsson}, \citenamefont {Castro~Neto}, \citenamefont {Guinea},\ and\
  \citenamefont {Peres}}]{nilssonTransmissionBiasedGraphene2007}%
  \BibitemOpen
  \bibfield  {author} {\bibinfo {author} {\bibfnamefont {J.}~\bibnamefont
  {Nilsson}}, \bibinfo {author} {\bibfnamefont {A.~H.}\ \bibnamefont
  {Castro~Neto}}, \bibinfo {author} {\bibfnamefont {F.}~\bibnamefont
  {Guinea}},\ and\ \bibinfo {author} {\bibfnamefont {N.~M.~R.}\ \bibnamefont
  {Peres}},\ }\bibfield  {title} {\bibinfo {title} {Transmission through a
  biased graphene bilayer barrier},\ }\href
  {https://doi.org/10.1103/PhysRevB.76.165416} {\bibfield  {journal} {\bibinfo
  {journal} {Physical Review B}\ }\textbf {\bibinfo {volume} {76}},\ \bibinfo
  {pages} {165416} (\bibinfo {year} {2007})}\BibitemShut {NoStop}%
\bibitem [{\citenamefont {Nakanishi}\ \emph {et~al.}(2011)\citenamefont
  {Nakanishi}, \citenamefont {Koshino},\ and\ \citenamefont
  {Ando}}]{nakanishiRoleEvanescentWave2011}%
  \BibitemOpen
  \bibfield  {author} {\bibinfo {author} {\bibfnamefont {T.}~\bibnamefont
  {Nakanishi}}, \bibinfo {author} {\bibfnamefont {M.}~\bibnamefont {Koshino}},\
  and\ \bibinfo {author} {\bibfnamefont {T.}~\bibnamefont {Ando}},\ }\bibfield
  {title} {\bibinfo {title} {Role of evanescent wave in valley polarization
  through junction of mono- and bi-layer graphenes},\ }\href
  {https://doi.org/10.1088/1742-6596/302/1/012021} {\bibfield  {journal}
  {\bibinfo  {journal} {Journal of Physics: Conference Series}\ }\textbf
  {\bibinfo {volume} {302}},\ \bibinfo {pages} {012021} (\bibinfo {year}
  {2011})}\BibitemShut {NoStop}%
\bibitem [{\citenamefont {Milovanovi{\'c}}\ \emph {et~al.}(2013)\citenamefont
  {Milovanovi{\'c}}, \citenamefont {Ramezani~Masir},\ and\ \citenamefont
  {Peeters}}]{milovanovicBilayerGrapheneHall2013}%
  \BibitemOpen
  \bibfield  {author} {\bibinfo {author} {\bibfnamefont {S.~P.}\ \bibnamefont
  {Milovanovi{\'c}}}, \bibinfo {author} {\bibfnamefont {M.}~\bibnamefont
  {Ramezani~Masir}},\ and\ \bibinfo {author} {\bibfnamefont {F.~M.}\
  \bibnamefont {Peeters}},\ }\bibfield  {title} {\bibinfo {title} {Bilayer
  graphene {{Hall}} bar with a pn-junction},\ }\href
  {https://doi.org/10.1063/1.4821264} {\bibfield  {journal} {\bibinfo
  {journal} {Journal of Applied Physics}\ }\textbf {\bibinfo {volume} {114}},\
  \bibinfo {pages} {113706} (\bibinfo {year} {2013})}\BibitemShut {NoStop}%
\bibitem [{\citenamefont {Park}(2014)}]{parkBandGapTunedOscillatory2014}%
  \BibitemOpen
  \bibfield  {author} {\bibinfo {author} {\bibfnamefont {C.-S.}\ \bibnamefont
  {Park}},\ }\bibfield  {title} {\bibinfo {title} {Band-{{Gap}} tuned
  oscillatory conductance in bilayer graphene n-p-n junction},\ }\href
  {https://doi.org/10.1063/1.4890224} {\bibfield  {journal} {\bibinfo
  {journal} {Journal of Applied Physics}\ }\textbf {\bibinfo {volume} {116}},\
  \bibinfo {pages} {033702} (\bibinfo {year} {2014})}\BibitemShut {NoStop}%
\bibitem [{\citenamefont {Nakanishi}\ \emph {et~al.}(2010)\citenamefont
  {Nakanishi}, \citenamefont {Koshino},\ and\ \citenamefont
  {Ando}}]{nakanishiTransmissionBoundaryMonolayer2010}%
  \BibitemOpen
  \bibfield  {author} {\bibinfo {author} {\bibfnamefont {T.}~\bibnamefont
  {Nakanishi}}, \bibinfo {author} {\bibfnamefont {M.}~\bibnamefont {Koshino}},\
  and\ \bibinfo {author} {\bibfnamefont {T.}~\bibnamefont {Ando}},\ }\bibfield
  {title} {\bibinfo {title} {Transmission through a boundary between monolayer
  and bilayer graphene},\ }\href {https://doi.org/10.1103/PhysRevB.82.125428}
  {\bibfield  {journal} {\bibinfo  {journal} {Physical Review B}\ }\textbf
  {\bibinfo {volume} {82}},\ \bibinfo {pages} {125428} (\bibinfo {year}
  {2010})}\BibitemShut {NoStop}%
\bibitem [{\citenamefont {Barbier}\ \emph {et~al.}(2010)\citenamefont
  {Barbier}, \citenamefont {Vasilopoulos},\ and\ \citenamefont
  {Peeters}}]{barbierKronigPenneyModelBilayer2010}%
  \BibitemOpen
  \bibfield  {author} {\bibinfo {author} {\bibfnamefont {M.}~\bibnamefont
  {Barbier}}, \bibinfo {author} {\bibfnamefont {P.}~\bibnamefont
  {Vasilopoulos}},\ and\ \bibinfo {author} {\bibfnamefont {F.~M.}\ \bibnamefont
  {Peeters}},\ }\bibfield  {title} {\bibinfo {title} {Kronig-{{Penney}} model
  on bilayer graphene: {{Spectrum}} and transmission periodic in the strength
  of the barriers},\ }\href {https://doi.org/10.1103/PhysRevB.82.235408}
  {\bibfield  {journal} {\bibinfo  {journal} {Physical Review B}\ }\textbf
  {\bibinfo {volume} {82}},\ \bibinfo {pages} {235408} (\bibinfo {year}
  {2010})}\BibitemShut {NoStop}%
\bibitem [{\citenamefont {Van~Duppen}\ and\ \citenamefont
  {Peeters}(2013)}]{vanduppenFourbandTunnelingBilayer2013}%
  \BibitemOpen
  \bibfield  {author} {\bibinfo {author} {\bibfnamefont {B.}~\bibnamefont
  {Van~Duppen}}\ and\ \bibinfo {author} {\bibfnamefont {F.~M.}\ \bibnamefont
  {Peeters}},\ }\bibfield  {title} {\bibinfo {title} {Four-band tunneling in
  bilayer graphene},\ }\href {https://doi.org/10.1103/PhysRevB.87.205427}
  {\bibfield  {journal} {\bibinfo  {journal} {Physical Review B}\ }\textbf
  {\bibinfo {volume} {87}},\ \bibinfo {pages} {205427} (\bibinfo {year}
  {2013})}\BibitemShut {NoStop}%
\bibitem [{\citenamefont {Sa{\~n}udo}\ and\ \citenamefont
  {{Lopez-Ruiz}}(2014)}]{sanudoStatisticalMagnitudesKlein2014}%
  \BibitemOpen
  \bibfield  {author} {\bibinfo {author} {\bibfnamefont {J.}~\bibnamefont
  {Sa{\~n}udo}}\ and\ \bibinfo {author} {\bibfnamefont {R.}~\bibnamefont
  {{Lopez-Ruiz}}},\ }\href {https://doi.org/10.48550/arXiv.1408.5333} {\bibinfo
  {title} {Statistical magnitudes and the {{Klein}} tunneling in bi-layer
  graphene: Influence of evanescent waves}} (\bibinfo {year} {2014}),\ \Eprint
  {https://arxiv.org/abs/1408.5333} {arXiv:1408.5333 [cond-mat, physics:nlin,
  physics:quant-ph]} \BibitemShut {NoStop}%
\bibitem [{\citenamefont {Saley}\ \emph {et~al.}(2022)\citenamefont {Saley},
  \citenamefont {Mouhafid}, \citenamefont {Jellal},\ and\ \citenamefont
  {Siari}}]{saleyKleinTunnelingTriple2022}%
  \BibitemOpen
  \bibfield  {author} {\bibinfo {author} {\bibfnamefont {M.~H.}\ \bibnamefont
  {Saley}}, \bibinfo {author} {\bibfnamefont {A.~E.}\ \bibnamefont {Mouhafid}},
  \bibinfo {author} {\bibfnamefont {A.}~\bibnamefont {Jellal}},\ and\ \bibinfo
  {author} {\bibfnamefont {A.}~\bibnamefont {Siari}},\ }\href
  {https://doi.org/10.48550/arXiv.2206.15446} {\bibinfo {title} {Klein
  tunneling through triple barrier in {{AB}} bilayer graphene}} (\bibinfo
  {year} {2022}),\ \Eprint {https://arxiv.org/abs/2206.15446} {arXiv:2206.15446
  [cond-mat]} \BibitemShut {NoStop}%
\bibitem [{\citenamefont {He}\ \emph {et~al.}(2013)\citenamefont {He},
  \citenamefont {Chu},\ and\ \citenamefont
  {He}}]{heChiralTunnelingTwisted2013}%
  \BibitemOpen
  \bibfield  {author} {\bibinfo {author} {\bibfnamefont {W.-Y.}\ \bibnamefont
  {He}}, \bibinfo {author} {\bibfnamefont {Z.-D.}\ \bibnamefont {Chu}},\ and\
  \bibinfo {author} {\bibfnamefont {L.}~\bibnamefont {He}},\ }\bibfield
  {title} {\bibinfo {title} {Chiral {{Tunneling}} in a {{Twisted Graphene
  Bilayer}}},\ }\href {https://doi.org/10.1103/PhysRevLett.111.066803}
  {\bibfield  {journal} {\bibinfo  {journal} {Physical Review Letters}\
  }\textbf {\bibinfo {volume} {111}},\ \bibinfo {pages} {066803} (\bibinfo
  {year} {2013})}\BibitemShut {NoStop}%
\bibitem [{\citenamefont {Stigloher}\ \emph {et~al.}(2016)\citenamefont
  {Stigloher}, \citenamefont {Decker}, \citenamefont {K{\"o}rner},
  \citenamefont {Tanabe}, \citenamefont {Moriyama}, \citenamefont {Taniguchi},
  \citenamefont {Hata}, \citenamefont {Madami}, \citenamefont {Gubbiotti},
  \citenamefont {Kobayashi}, \citenamefont {Ono},\ and\ \citenamefont
  {Back}}]{stigloherSnellLawSpin2016}%
  \BibitemOpen
  \bibfield  {author} {\bibinfo {author} {\bibfnamefont {J.}~\bibnamefont
  {Stigloher}}, \bibinfo {author} {\bibfnamefont {M.}~\bibnamefont {Decker}},
  \bibinfo {author} {\bibfnamefont {H.~S.}\ \bibnamefont {K{\"o}rner}},
  \bibinfo {author} {\bibfnamefont {K.}~\bibnamefont {Tanabe}}, \bibinfo
  {author} {\bibfnamefont {T.}~\bibnamefont {Moriyama}}, \bibinfo {author}
  {\bibfnamefont {T.}~\bibnamefont {Taniguchi}}, \bibinfo {author}
  {\bibfnamefont {H.}~\bibnamefont {Hata}}, \bibinfo {author} {\bibfnamefont
  {M.}~\bibnamefont {Madami}}, \bibinfo {author} {\bibfnamefont
  {G.}~\bibnamefont {Gubbiotti}}, \bibinfo {author} {\bibfnamefont
  {K.}~\bibnamefont {Kobayashi}}, \bibinfo {author} {\bibfnamefont
  {T.}~\bibnamefont {Ono}},\ and\ \bibinfo {author} {\bibfnamefont {C.~H.}\
  \bibnamefont {Back}},\ }\bibfield  {title} {\bibinfo {title} {Snell's {{Law}}
  for {{Spin Waves}}},\ }\href {https://doi.org/10.1103/PhysRevLett.117.037204}
  {\bibfield  {journal} {\bibinfo  {journal} {Physical Review Letters}\
  }\textbf {\bibinfo {volume} {117}},\ \bibinfo {pages} {037204} (\bibinfo
  {year} {2016})}\BibitemShut {NoStop}%
\bibitem [{\citenamefont {Ahn}\ and\ \citenamefont
  {Das~Sarma}(2021)}]{ahnTheoryAnisotropicPlasmons2021}%
  \BibitemOpen
  \bibfield  {author} {\bibinfo {author} {\bibfnamefont {S.}~\bibnamefont
  {Ahn}}\ and\ \bibinfo {author} {\bibfnamefont {S.}~\bibnamefont
  {Das~Sarma}},\ }\bibfield  {title} {\bibinfo {title} {Theory of anisotropic
  plasmons},\ }\href {https://doi.org/10.1103/PhysRevB.103.L041303} {\bibfield
  {journal} {\bibinfo  {journal} {Physical Review B}\ }\textbf {\bibinfo
  {volume} {103}},\ \bibinfo {pages} {L041303} (\bibinfo {year}
  {2021})}\BibitemShut {NoStop}%
\bibitem [{\citenamefont {Alagappan}\ \emph {et~al.}(2006)\citenamefont
  {Alagappan}, \citenamefont {Sun}, \citenamefont {Shum}, \citenamefont {Yu},\
  and\ \citenamefont {den
  Engelsen}}]{alagappanSymmetryPropertiesTwodimensional2006}%
  \BibitemOpen
  \bibfield  {author} {\bibinfo {author} {\bibfnamefont {G.}~\bibnamefont
  {Alagappan}}, \bibinfo {author} {\bibfnamefont {X.~W.}\ \bibnamefont {Sun}},
  \bibinfo {author} {\bibfnamefont {P.}~\bibnamefont {Shum}}, \bibinfo {author}
  {\bibfnamefont {M.~B.}\ \bibnamefont {Yu}},\ and\ \bibinfo {author}
  {\bibfnamefont {D.}~\bibnamefont {den Engelsen}},\ }\bibfield  {title}
  {\bibinfo {title} {Symmetry properties of two-dimensional anisotropic
  photonic crystals},\ }\href {https://doi.org/10.1364/JOSAA.23.002002}
  {\bibfield  {journal} {\bibinfo  {journal} {JOSA A}\ }\textbf {\bibinfo
  {volume} {23}},\ \bibinfo {pages} {2002} (\bibinfo {year}
  {2006})}\BibitemShut {NoStop}%
\bibitem [{\citenamefont {Khromova}\ and\ \citenamefont
  {Melnikov}(2008)}]{khromovaAnisotropicPhotonicCrystals2008}%
  \BibitemOpen
  \bibfield  {author} {\bibinfo {author} {\bibfnamefont {I.~A.}\ \bibnamefont
  {Khromova}}\ and\ \bibinfo {author} {\bibfnamefont {L.~A.}\ \bibnamefont
  {Melnikov}},\ }\bibfield  {title} {\bibinfo {title} {Anisotropic photonic
  crystals: {{Generalized}} plane wave method and dispersion symmetry
  properties},\ }\href {https://doi.org/10.1016/j.optcom.2008.07.059}
  {\bibfield  {journal} {\bibinfo  {journal} {Optics Communications}\ }\textbf
  {\bibinfo {volume} {281}},\ \bibinfo {pages} {5458} (\bibinfo {year}
  {2008})}\BibitemShut {NoStop}%
\bibitem [{\citenamefont {P{\'e}terfalvi}\ \emph {et~al.}(2012)\citenamefont
  {P{\'e}terfalvi}, \citenamefont {Oroszl{\'a}ny}, \citenamefont {Lambert},\
  and\ \citenamefont {Cserti}}]{peterfalviIntrabandElectronFocusing2012}%
  \BibitemOpen
  \bibfield  {author} {\bibinfo {author} {\bibfnamefont {C.~G.}\ \bibnamefont
  {P{\'e}terfalvi}}, \bibinfo {author} {\bibfnamefont {L.}~\bibnamefont
  {Oroszl{\'a}ny}}, \bibinfo {author} {\bibfnamefont {C.~J.}\ \bibnamefont
  {Lambert}},\ and\ \bibinfo {author} {\bibfnamefont {J.}~\bibnamefont
  {Cserti}},\ }\bibfield  {title} {\bibinfo {title} {Intraband electron
  focusing in bilayer graphene},\ }\href
  {https://doi.org/10.1088/1367-2630/14/6/063028} {\bibfield  {journal}
  {\bibinfo  {journal} {New Journal of Physics}\ }\textbf {\bibinfo {volume}
  {14}},\ \bibinfo {pages} {063028} (\bibinfo {year} {2012})}\BibitemShut
  {NoStop}%
\bibitem [{Note1()}]{Note1}%
  \BibitemOpen
  \bibinfo {note} {Note that $k_{\parallel }$ replaces the commonly used sine
  of the angle of incidence as momentum variable \cite
  {berryRegularityChaosClassical1981, meissSymplecticMapsVariational1992,
  meissCantoriStadiumBilliard1992} since it is the conserved momentum in the
  present anisotropic situation, cf.~the Supplemental Material for more
  details.}\BibitemShut {Stop}%
\bibitem [{\citenamefont {McCann}\ and\ \citenamefont
  {Fal'ko}(2006)}]{mccannLandauLevelDegeneracyQuantum2006}%
  \BibitemOpen
  \bibfield  {author} {\bibinfo {author} {\bibfnamefont {E.}~\bibnamefont
  {McCann}}\ and\ \bibinfo {author} {\bibfnamefont {V.~I.}\ \bibnamefont
  {Fal'ko}},\ }\bibfield  {title} {\bibinfo {title} {Landau-{{Level
  Degeneracy}} and {{Quantum Hall Effect}} in a {{Graphite Bilayer}}},\ }\href
  {https://doi.org/10.1103/PhysRevLett.96.086805} {\bibfield  {journal}
  {\bibinfo  {journal} {Physical Review Letters}\ }\textbf {\bibinfo {volume}
  {96}},\ \bibinfo {pages} {086805} (\bibinfo {year} {2006})}\BibitemShut
  {NoStop}%
\bibitem [{\citenamefont {McCann}\ \emph {et~al.}(2007)\citenamefont {McCann},
  \citenamefont {Abergel},\ and\ \citenamefont
  {Fal'ko}}]{mccannLowEnergyElectronic2007}%
  \BibitemOpen
  \bibfield  {author} {\bibinfo {author} {\bibfnamefont {E.}~\bibnamefont
  {McCann}}, \bibinfo {author} {\bibfnamefont {D.~S.}\ \bibnamefont
  {Abergel}},\ and\ \bibinfo {author} {\bibfnamefont {V.~I.}\ \bibnamefont
  {Fal'ko}},\ }\bibfield  {title} {\bibinfo {title} {The low energy electronic
  band structure of bilayer graphene},\ }\href
  {https://doi.org/10.1140/epjst/e2007-00229-1} {\bibfield  {journal} {\bibinfo
   {journal} {The European Physical Journal Special Topics}\ }\textbf {\bibinfo
  {volume} {148}},\ \bibinfo {pages} {91} (\bibinfo {year} {2007})}\BibitemShut
  {NoStop}%
\bibitem [{\citenamefont {McCann}\ and\ \citenamefont
  {Koshino}(2013)}]{mccannElectronicPropertiesBilayer2013}%
  \BibitemOpen
  \bibfield  {author} {\bibinfo {author} {\bibfnamefont {E.}~\bibnamefont
  {McCann}}\ and\ \bibinfo {author} {\bibfnamefont {M.}~\bibnamefont
  {Koshino}},\ }\bibfield  {title} {\bibinfo {title} {The electronic properties
  of bilayer graphene},\ }\href {https://doi.org/10.1088/0034-4885/76/5/056503}
  {\bibfield  {journal} {\bibinfo  {journal} {Reports on Progress in Physics}\
  }\textbf {\bibinfo {volume} {76}},\ \bibinfo {pages} {056503} (\bibinfo
  {year} {2013})}\BibitemShut {NoStop}%
\bibitem [{\citenamefont {Kuzmenko}\ \emph {et~al.}(2009)\citenamefont
  {Kuzmenko}, \citenamefont {Crassee}, \citenamefont {{van der Marel}},
  \citenamefont {Blake},\ and\ \citenamefont
  {Novoselov}}]{kuzmenkoDeterminationGatetunableBand2009}%
  \BibitemOpen
  \bibfield  {author} {\bibinfo {author} {\bibfnamefont {A.~B.}\ \bibnamefont
  {Kuzmenko}}, \bibinfo {author} {\bibfnamefont {I.}~\bibnamefont {Crassee}},
  \bibinfo {author} {\bibfnamefont {D.}~\bibnamefont {{van der Marel}}},
  \bibinfo {author} {\bibfnamefont {P.}~\bibnamefont {Blake}},\ and\ \bibinfo
  {author} {\bibfnamefont {K.~S.}\ \bibnamefont {Novoselov}},\ }\bibfield
  {title} {\bibinfo {title} {Determination of the gate-tunable band gap and
  tight-binding parameters in bilayer graphene using infrared spectroscopy},\
  }\href {https://doi.org/10.1103/PhysRevB.80.165406} {\bibfield  {journal}
  {\bibinfo  {journal} {Physical Review B}\ }\textbf {\bibinfo {volume} {80}},\
  \bibinfo {pages} {165406} (\bibinfo {year} {2009})}\BibitemShut {NoStop}%
\bibitem [{\citenamefont {Kraft}\ \emph {et~al.}(2020)\citenamefont {Kraft},
  \citenamefont {Liu}, \citenamefont {Selvasundaram}, \citenamefont {Chen},
  \citenamefont {Krupke}, \citenamefont {Richter},\ and\ \citenamefont
  {Danneau}}]{kraftAnomalousCyclotronMotion2020}%
  \BibitemOpen
  \bibfield  {author} {\bibinfo {author} {\bibfnamefont {R.}~\bibnamefont
  {Kraft}}, \bibinfo {author} {\bibfnamefont {M.-H.}\ \bibnamefont {Liu}},
  \bibinfo {author} {\bibfnamefont {P.~B.}\ \bibnamefont {Selvasundaram}},
  \bibinfo {author} {\bibfnamefont {S.-C.}\ \bibnamefont {Chen}}, \bibinfo
  {author} {\bibfnamefont {R.}~\bibnamefont {Krupke}}, \bibinfo {author}
  {\bibfnamefont {K.}~\bibnamefont {Richter}},\ and\ \bibinfo {author}
  {\bibfnamefont {R.}~\bibnamefont {Danneau}},\ }\bibfield  {title} {\bibinfo
  {title} {Anomalous {{Cyclotron Motion}} in {{Graphene Superlattice
  Cavities}}},\ }\href {https://doi.org/10.1103/PhysRevLett.125.217701}
  {\bibfield  {journal} {\bibinfo  {journal} {Physical Review Letters}\
  }\textbf {\bibinfo {volume} {125}},\ \bibinfo {pages} {217701} (\bibinfo
  {year} {2020})}\BibitemShut {NoStop}%
\bibitem [{Note2()}]{Note2}%
  \BibitemOpen
  \bibinfo {note} {Generalizations to smooth confinement potentials have been
  discussed, e.g., in Refs.~\protect \rev@citealp
  {cheianovSelectiveTransmissionDirac2006,
  peterfalviIntrabandElectronFocusing2012, schrepferDiracFermionOptics2021}.
  See the conclusion for further discussion.}\BibitemShut {Stop}%
\bibitem [{Note3()}]{Note3}%
  \BibitemOpen
  \bibinfo {note} {We neglect intervalley scattering and treat the two valleys
  separately. This is justified due to the high quality of current bilayer
  graphene{} samples not hosting any atomic-scale defects or abrupt lattice
  termination \cite {goldCoherentJettingGateDefined2021}}\BibitemShut {NoStop}%
\bibitem [{\citenamefont {Berry}(1981)}]{berryRegularityChaosClassical1981}%
  \BibitemOpen
  \bibfield  {author} {\bibinfo {author} {\bibfnamefont {M.~V.}\ \bibnamefont
  {Berry}},\ }\bibfield  {title} {\bibinfo {title} {Regularity and chaos in
  classical mechanics, illustrated by three deformations of a circular
  'billiard'},\ }\href {https://doi.org/10.1088/0143-0807/2/2/006} {\bibfield
  {journal} {\bibinfo  {journal} {European Journal of Physics}\ }\textbf
  {\bibinfo {volume} {2}},\ \bibinfo {pages} {91} (\bibinfo {year}
  {1981})}\BibitemShut {NoStop}%
\bibitem [{\citenamefont {Meiss}(1992)}]{meissCantoriStadiumBilliard1992}%
  \BibitemOpen
  \bibfield  {author} {\bibinfo {author} {\bibfnamefont {J.~D.}\ \bibnamefont
  {Meiss}},\ }\bibfield  {title} {\bibinfo {title} {Cantori for the stadium
  billiard},\ }\href {https://doi.org/10.1063/1.165867} {\bibfield  {journal}
  {\bibinfo  {journal} {Chaos: An Interdisciplinary Journal of Nonlinear
  Science}\ }\textbf {\bibinfo {volume} {2}},\ \bibinfo {pages} {267} (\bibinfo
  {year} {1992})}\BibitemShut {NoStop}%
\bibitem [{\citenamefont {Xiao}\ \emph {et~al.}(2010)\citenamefont {Xiao},
  \citenamefont {Zou}, \citenamefont {Li}, \citenamefont {Dong}, \citenamefont
  {Han},\ and\ \citenamefont {Gong}}]{xiaoAsymmetricResonantCavities2010}%
  \BibitemOpen
  \bibfield  {author} {\bibinfo {author} {\bibfnamefont {Y.-F.}\ \bibnamefont
  {Xiao}}, \bibinfo {author} {\bibfnamefont {C.-L.}\ \bibnamefont {Zou}},
  \bibinfo {author} {\bibfnamefont {Y.}~\bibnamefont {Li}}, \bibinfo {author}
  {\bibfnamefont {C.-H.}\ \bibnamefont {Dong}}, \bibinfo {author}
  {\bibfnamefont {Z.-F.}\ \bibnamefont {Han}},\ and\ \bibinfo {author}
  {\bibfnamefont {Q.}~\bibnamefont {Gong}},\ }\bibfield  {title} {\bibinfo
  {title} {Asymmetric resonant cavities and their applications in optics and
  photonics: A review},\ }\href {https://doi.org/10.1007/s12200-010-0003-2}
  {\bibfield  {journal} {\bibinfo  {journal} {Frontiers of Optoelectronics in
  China}\ }\textbf {\bibinfo {volume} {3}},\ \bibinfo {pages} {109} (\bibinfo
  {year} {2010})}\BibitemShut {NoStop}%
\bibitem [{\citenamefont {Wiersig}\ and\ \citenamefont
  {Hentschel}(2008)}]{wiersigCombiningDirectionalLight2008}%
  \BibitemOpen
  \bibfield  {author} {\bibinfo {author} {\bibfnamefont {J.}~\bibnamefont
  {Wiersig}}\ and\ \bibinfo {author} {\bibfnamefont {M.}~\bibnamefont
  {Hentschel}},\ }\bibfield  {title} {\bibinfo {title} {Combining {{Directional
  Light Output}} and {{Ultralow Loss}} in {{Deformed Microdisks}}},\ }\href
  {https://doi.org/10.1103/PhysRevLett.100.033901} {\bibfield  {journal}
  {\bibinfo  {journal} {Physical Review Letters}\ }\textbf {\bibinfo {volume}
  {100}},\ \bibinfo {pages} {033901} (\bibinfo {year} {2008})}\BibitemShut
  {NoStop}%
\bibitem [{\citenamefont {Cheianov}\ and\ \citenamefont
  {Fal'ko}(2006)}]{cheianovSelectiveTransmissionDirac2006}%
  \BibitemOpen
  \bibfield  {author} {\bibinfo {author} {\bibfnamefont {V.~V.}\ \bibnamefont
  {Cheianov}}\ and\ \bibinfo {author} {\bibfnamefont {V.~I.}\ \bibnamefont
  {Fal'ko}},\ }\bibfield  {title} {\bibinfo {title} {Selective transmission of
  {{Dirac}} electrons and ballistic magnetoresistance of n - p junctions in
  graphene},\ }\href {https://doi.org/10.1103/PhysRevB.74.041403} {\bibfield
  {journal} {\bibinfo  {journal} {Physical Review B}\ }\textbf {\bibinfo
  {volume} {74}},\ \bibinfo {pages} {041403} (\bibinfo {year}
  {2006})}\BibitemShut {NoStop}%
\bibitem [{\citenamefont {Varlet}\ \emph {et~al.}(2015)\citenamefont {Varlet},
  \citenamefont {{Mucha-Kruczy{\'n}ski}}, \citenamefont {Bischoff},
  \citenamefont {Simonet}, \citenamefont {Taniguchi}, \citenamefont {Watanabe},
  \citenamefont {Fal'ko}, \citenamefont {Ihn},\ and\ \citenamefont
  {Ensslin}}]{varletTunableFermiSurface2015}%
  \BibitemOpen
  \bibfield  {author} {\bibinfo {author} {\bibfnamefont {A.}~\bibnamefont
  {Varlet}}, \bibinfo {author} {\bibfnamefont {M.}~\bibnamefont
  {{Mucha-Kruczy{\'n}ski}}}, \bibinfo {author} {\bibfnamefont {D.}~\bibnamefont
  {Bischoff}}, \bibinfo {author} {\bibfnamefont {P.}~\bibnamefont {Simonet}},
  \bibinfo {author} {\bibfnamefont {T.}~\bibnamefont {Taniguchi}}, \bibinfo
  {author} {\bibfnamefont {K.}~\bibnamefont {Watanabe}}, \bibinfo {author}
  {\bibfnamefont {V.}~\bibnamefont {Fal'ko}}, \bibinfo {author} {\bibfnamefont
  {T.}~\bibnamefont {Ihn}},\ and\ \bibinfo {author} {\bibfnamefont
  {K.}~\bibnamefont {Ensslin}},\ }\bibfield  {title} {\bibinfo {title} {Tunable
  {{Fermi}} surface topology and {{Lifshitz}} transition in bilayer graphene},\
  }\href {https://doi.org/10.1016/j.synthmet.2015.07.006} {\bibfield  {journal}
  {\bibinfo  {journal} {Synthetic Metals}\ }\bibinfo {series} {Reviews of
  {{Current Advances}} in {{Graphene Science}} and {{Technology}}},\ \textbf
  {\bibinfo {volume} {210}},\ \bibinfo {pages} {19} (\bibinfo {year}
  {2015})}\BibitemShut {NoStop}%
\bibitem [{\citenamefont {Varlet}\ \emph {et~al.}(2014)\citenamefont {Varlet},
  \citenamefont {Bischoff}, \citenamefont {Simonet}, \citenamefont {Watanabe},
  \citenamefont {Taniguchi}, \citenamefont {Ihn}, \citenamefont {Ensslin},
  \citenamefont {{Mucha-Kruczy{\'n}ski}},\ and\ \citenamefont
  {Fal'ko}}]{varletAnomalousSequenceQuantum2014}%
  \BibitemOpen
  \bibfield  {author} {\bibinfo {author} {\bibfnamefont {A.}~\bibnamefont
  {Varlet}}, \bibinfo {author} {\bibfnamefont {D.}~\bibnamefont {Bischoff}},
  \bibinfo {author} {\bibfnamefont {P.}~\bibnamefont {Simonet}}, \bibinfo
  {author} {\bibfnamefont {K.}~\bibnamefont {Watanabe}}, \bibinfo {author}
  {\bibfnamefont {T.}~\bibnamefont {Taniguchi}}, \bibinfo {author}
  {\bibfnamefont {T.}~\bibnamefont {Ihn}}, \bibinfo {author} {\bibfnamefont
  {K.}~\bibnamefont {Ensslin}}, \bibinfo {author} {\bibfnamefont
  {M.}~\bibnamefont {{Mucha-Kruczy{\'n}ski}}},\ and\ \bibinfo {author}
  {\bibfnamefont {V.~I.}\ \bibnamefont {Fal'ko}},\ }\bibfield  {title}
  {\bibinfo {title} {Anomalous {{Sequence}} of {{Quantum Hall Liquids
  Revealing}} a {{Tunable Lifshitz Transition}} in {{Bilayer Graphene}}},\
  }\href {https://doi.org/10.1103/PhysRevLett.113.116602} {\bibfield  {journal}
  {\bibinfo  {journal} {Physical Review Letters}\ }\textbf {\bibinfo {volume}
  {113}},\ \bibinfo {pages} {116602} (\bibinfo {year} {2014})}\BibitemShut
  {NoStop}%
\bibitem [{\citenamefont {Moulsdale}\ \emph {et~al.}(2020)\citenamefont
  {Moulsdale}, \citenamefont {Knothe},\ and\ \citenamefont
  {Fal'ko}}]{moulsdaleEngineeringTopologicalMagnetic2020}%
  \BibitemOpen
  \bibfield  {author} {\bibinfo {author} {\bibfnamefont {C.}~\bibnamefont
  {Moulsdale}}, \bibinfo {author} {\bibfnamefont {A.}~\bibnamefont {Knothe}},\
  and\ \bibinfo {author} {\bibfnamefont {V.}~\bibnamefont {Fal'ko}},\
  }\bibfield  {title} {\bibinfo {title} {Engineering of the topological
  magnetic moment of electrons in bilayer graphene using strain and electrical
  bias},\ }\href {https://doi.org/10.1103/PhysRevB.101.085118} {\bibfield
  {journal} {\bibinfo  {journal} {Physical Review B}\ }\textbf {\bibinfo
  {volume} {101}},\ \bibinfo {pages} {085118} (\bibinfo {year}
  {2020})}\BibitemShut {NoStop}%
\bibitem [{\citenamefont {Morikawa}\ \emph {et~al.}(2015)\citenamefont
  {Morikawa}, \citenamefont {Dou}, \citenamefont {Wang}, \citenamefont {Smith},
  \citenamefont {Watanabe}, \citenamefont {Taniguchi}, \citenamefont
  {Masubuchi}, \citenamefont {Machida},\ and\ \citenamefont
  {Connolly}}]{morikawaImagingBallisticCarrier2015}%
  \BibitemOpen
  \bibfield  {author} {\bibinfo {author} {\bibfnamefont {S.}~\bibnamefont
  {Morikawa}}, \bibinfo {author} {\bibfnamefont {Z.}~\bibnamefont {Dou}},
  \bibinfo {author} {\bibfnamefont {S.-W.}\ \bibnamefont {Wang}}, \bibinfo
  {author} {\bibfnamefont {C.~G.}\ \bibnamefont {Smith}}, \bibinfo {author}
  {\bibfnamefont {K.}~\bibnamefont {Watanabe}}, \bibinfo {author}
  {\bibfnamefont {T.}~\bibnamefont {Taniguchi}}, \bibinfo {author}
  {\bibfnamefont {S.}~\bibnamefont {Masubuchi}}, \bibinfo {author}
  {\bibfnamefont {T.}~\bibnamefont {Machida}},\ and\ \bibinfo {author}
  {\bibfnamefont {M.~R.}\ \bibnamefont {Connolly}},\ }\bibfield  {title}
  {\bibinfo {title} {Imaging ballistic carrier trajectories in graphene using
  scanning gate microscopy},\ }\href {https://doi.org/10.1063/1.4937473}
  {\bibfield  {journal} {\bibinfo  {journal} {Applied Physics Letters}\
  }\textbf {\bibinfo {volume} {107}},\ \bibinfo {pages} {243102} (\bibinfo
  {year} {2015})}\BibitemShut {NoStop}%
\bibitem [{\citenamefont {Bhandari}\ \emph {et~al.}(2020)\citenamefont
  {Bhandari}, \citenamefont {Kreidel}, \citenamefont {Kelser}, \citenamefont
  {Lee}, \citenamefont {Watanabe}, \citenamefont {Taniguchi}, \citenamefont
  {Kim},\ and\ \citenamefont {Westervelt}}]{bhandariImagingFlowHoles2020}%
  \BibitemOpen
  \bibfield  {author} {\bibinfo {author} {\bibfnamefont {S.}~\bibnamefont
  {Bhandari}}, \bibinfo {author} {\bibfnamefont {M.}~\bibnamefont {Kreidel}},
  \bibinfo {author} {\bibfnamefont {A.}~\bibnamefont {Kelser}}, \bibinfo
  {author} {\bibfnamefont {G.-H.}\ \bibnamefont {Lee}}, \bibinfo {author}
  {\bibfnamefont {K.}~\bibnamefont {Watanabe}}, \bibinfo {author}
  {\bibfnamefont {T.}~\bibnamefont {Taniguchi}}, \bibinfo {author}
  {\bibfnamefont {P.}~\bibnamefont {Kim}},\ and\ \bibinfo {author}
  {\bibfnamefont {R.~M.}\ \bibnamefont {Westervelt}},\ }\bibfield  {title}
  {\bibinfo {title} {Imaging the flow of holes from a collimating contact in
  graphene},\ }\href {https://doi.org/10.1088/1361-6641/aba08d} {\bibfield
  {journal} {\bibinfo  {journal} {Semiconductor Science and Technology}\
  }\textbf {\bibinfo {volume} {35}},\ \bibinfo {pages} {09LT02} (\bibinfo
  {year} {2020})}\BibitemShut {NoStop}%
\bibitem [{\citenamefont {Steinacher}\ \emph {et~al.}(2018)\citenamefont
  {Steinacher}, \citenamefont {P{\"o}ltl}, \citenamefont {Kr{\"a}henmann},
  \citenamefont {Hofmann}, \citenamefont {Reichl}, \citenamefont {Zwerger},
  \citenamefont {Wegscheider}, \citenamefont {Jalabert}, \citenamefont
  {Ensslin}, \citenamefont {Weinmann},\ and\ \citenamefont
  {Ihn}}]{steinacherScanningGateExperiments2018}%
  \BibitemOpen
  \bibfield  {author} {\bibinfo {author} {\bibfnamefont {R.}~\bibnamefont
  {Steinacher}}, \bibinfo {author} {\bibfnamefont {C.}~\bibnamefont
  {P{\"o}ltl}}, \bibinfo {author} {\bibfnamefont {T.}~\bibnamefont
  {Kr{\"a}henmann}}, \bibinfo {author} {\bibfnamefont {A.}~\bibnamefont
  {Hofmann}}, \bibinfo {author} {\bibfnamefont {C.}~\bibnamefont {Reichl}},
  \bibinfo {author} {\bibfnamefont {W.}~\bibnamefont {Zwerger}}, \bibinfo
  {author} {\bibfnamefont {W.}~\bibnamefont {Wegscheider}}, \bibinfo {author}
  {\bibfnamefont {R.~A.}\ \bibnamefont {Jalabert}}, \bibinfo {author}
  {\bibfnamefont {K.}~\bibnamefont {Ensslin}}, \bibinfo {author} {\bibfnamefont
  {D.}~\bibnamefont {Weinmann}},\ and\ \bibinfo {author} {\bibfnamefont
  {T.}~\bibnamefont {Ihn}},\ }\bibfield  {title} {\bibinfo {title} {Scanning
  gate experiments: {{From}} strongly to weakly invasive probes},\ }\href
  {https://doi.org/10.1103/PhysRevB.98.075426} {\bibfield  {journal} {\bibinfo
  {journal} {Physical Review B}\ }\textbf {\bibinfo {volume} {98}},\ \bibinfo
  {pages} {075426} (\bibinfo {year} {2018})}\BibitemShut {NoStop}%
\bibitem [{\citenamefont {Chen}\ \emph {et~al.}(2022)\citenamefont {Chen},
  \citenamefont {Weick}, \citenamefont {Weinmann},\ and\ \citenamefont
  {Jalabert}}]{chenScanningGateMicroscopy2022}%
  \BibitemOpen
  \bibfield  {author} {\bibinfo {author} {\bibfnamefont {X.}~\bibnamefont
  {Chen}}, \bibinfo {author} {\bibfnamefont {G.}~\bibnamefont {Weick}},
  \bibinfo {author} {\bibfnamefont {D.}~\bibnamefont {Weinmann}},\ and\
  \bibinfo {author} {\bibfnamefont {R.~A.}\ \bibnamefont {Jalabert}},\ }\href
  {https://doi.org/10.48550/arXiv.2211.14004} {\bibinfo {title} {Scanning gate
  microscopy in graphene nanostructures}} (\bibinfo {year} {2022}),\ \Eprint
  {https://arxiv.org/abs/2211.14004} {arXiv:2211.14004 [cond-mat]} \BibitemShut
  {NoStop}%
\bibitem [{\citenamefont {Gold}\ \emph
  {et~al.}(2021{\natexlab{b}})\citenamefont {Gold}, \citenamefont {Br{\"a}m},
  \citenamefont {Ferguson}, \citenamefont {Kr{\"a}henmann}, \citenamefont
  {Hofmann}, \citenamefont {Steinacher}, \citenamefont {Fratus}, \citenamefont
  {Reichl}, \citenamefont {Wegscheider}, \citenamefont {Weinmann},
  \citenamefont {Ensslin},\ and\ \citenamefont
  {Ihn}}]{goldImagingSignaturesLocal2021}%
  \BibitemOpen
  \bibfield  {author} {\bibinfo {author} {\bibfnamefont {C.}~\bibnamefont
  {Gold}}, \bibinfo {author} {\bibfnamefont {B.~A.}\ \bibnamefont {Br{\"a}m}},
  \bibinfo {author} {\bibfnamefont {M.~S.}\ \bibnamefont {Ferguson}}, \bibinfo
  {author} {\bibfnamefont {T.}~\bibnamefont {Kr{\"a}henmann}}, \bibinfo
  {author} {\bibfnamefont {A.}~\bibnamefont {Hofmann}}, \bibinfo {author}
  {\bibfnamefont {R.}~\bibnamefont {Steinacher}}, \bibinfo {author}
  {\bibfnamefont {K.~R.}\ \bibnamefont {Fratus}}, \bibinfo {author}
  {\bibfnamefont {C.}~\bibnamefont {Reichl}}, \bibinfo {author} {\bibfnamefont
  {W.}~\bibnamefont {Wegscheider}}, \bibinfo {author} {\bibfnamefont
  {D.}~\bibnamefont {Weinmann}}, \bibinfo {author} {\bibfnamefont
  {K.}~\bibnamefont {Ensslin}},\ and\ \bibinfo {author} {\bibfnamefont
  {T.}~\bibnamefont {Ihn}},\ }\bibfield  {title} {\bibinfo {title} {Imaging
  signatures of the local density of states in an electronic cavity},\ }\href
  {https://doi.org/10.1103/PhysRevResearch.3.L032005} {\bibfield  {journal}
  {\bibinfo  {journal} {Physical Review Research}\ }\textbf {\bibinfo {volume}
  {3}},\ \bibinfo {pages} {L032005} (\bibinfo {year}
  {2021}{\natexlab{b}})}\BibitemShut {NoStop}%
\bibitem [{\citenamefont {Fratus}\ \emph {et~al.}(2021)\citenamefont {Fratus},
  \citenamefont {Le~Calonnec}, \citenamefont {Jalabert}, \citenamefont
  {Weick},\ and\ \citenamefont
  {Weinmann}}]{fratusSignaturesFoldedBranches2021}%
  \BibitemOpen
  \bibfield  {author} {\bibinfo {author} {\bibfnamefont {K.~R.}\ \bibnamefont
  {Fratus}}, \bibinfo {author} {\bibfnamefont {C.}~\bibnamefont {Le~Calonnec}},
  \bibinfo {author} {\bibfnamefont {R.}~\bibnamefont {Jalabert}}, \bibinfo
  {author} {\bibfnamefont {G.}~\bibnamefont {Weick}},\ and\ \bibinfo {author}
  {\bibfnamefont {D.}~\bibnamefont {Weinmann}},\ }\bibfield  {title} {\bibinfo
  {title} {Signatures of folded branches in the scanning gate microscopy of
  ballistic electronic cavities},\ }\href
  {https://doi.org/10.21468/SciPostPhys.10.3.069} {\bibfield  {journal}
  {\bibinfo  {journal} {SciPost Physics}\ }\textbf {\bibinfo {volume} {10}},\
  \bibinfo {pages} {069} (\bibinfo {year} {2021})}\BibitemShut {NoStop}%
\bibitem [{Note4()}]{Note4}%
  \BibitemOpen
  \bibinfo {note} {Here, we do not take into account the possible distribution
  of states in an actual channel attached to the cavity, as, e.g., in \protect
  \rev@citealp {goldCoherentJettingGateDefined2021}. Doing so is a
  straightforward extension of our work to describe any actual experimental
  setup.}\BibitemShut {Stop}%
\bibitem [{\citenamefont {Knothe}\ and\ \citenamefont
  {Fal'ko}(2018)}]{knotheInfluenceMinivalleysBerry2018}%
  \BibitemOpen
  \bibfield  {author} {\bibinfo {author} {\bibfnamefont {A.}~\bibnamefont
  {Knothe}}\ and\ \bibinfo {author} {\bibfnamefont {V.}~\bibnamefont
  {Fal'ko}},\ }\bibfield  {title} {\bibinfo {title} {Influence of minivalleys
  and {{Berry}} curvature on electrostatically induced quantum wires in gapped
  bilayer graphene},\ }\href {https://doi.org/10.1103/PhysRevB.98.155435}
  {\bibfield  {journal} {\bibinfo  {journal} {Physical Review B}\ }\textbf
  {\bibinfo {volume} {98}},\ \bibinfo {pages} {155435} (\bibinfo {year}
  {2018})}\BibitemShut {NoStop}%
\end{thebibliography}%


\renewcommand\thefigure{S.\arabic{figure}} 
\renewcommand\theequation{S.\arabic{equation}}  
\setcounter{figure}{0} 
\setcounter{equation}{0}
\section*{Supplemantary Material}

 \section{Billiards dynamic simulation}
  Based on particle-wave correspondence, ballistic electrons can be followed by trajectory tracing through the cavity. Whereas this is a straightforward task for generic billiards with hard walls and even for billiards for light where the angles of incidence and reflection are equal, and the next reflection point is determined as the intersection of the trajectory with the billiard boundary, here things are more involved due to the anisotropic material properties induced by trigonal warping of the Fermi line. 
  
  
  We start with the four band Hamiltonian that we repeat here
  \begin{equation}
   H^{\xi}_{BLG}=\xi\left(\begin{matrix}
    \xi U-\frac{1}{2}\Delta  & v_3\pi & v_4 \pi^{\dagger} &v \pi^{\dagger}\\
    v_3 \pi^{\dagger}&\xi U+\frac{1}{2}\Delta & v\pi &v_4 \pi\\
    v_4 \pi & v\pi^{\dagger} & \xi U+\frac{1}{2}\Delta  & \xi \gamma_1\\
    v\pi & v_4 \pi^{\dagger} & \xi \gamma_1 &\xi U-\frac{1}{2}\Delta 
    \label{eq1}
   \end{matrix}\right)
  \end{equation}    
  using $\pi=p_x+ip_y$ and $p=\hbar k$.
    The trajectory tracing algorithm follows five steps, thereby changing between real and momentum space:
  \begin{enumerate}
   \item[A.] Set the initial condition by choosing a starting point and a first angle of incidence $\chi_{\mathrm{in}}$.  
   %
    \item[B.] Find the next intersection (or scattering) 
point, of the trajectory with the
boundary of the cavity, cf.~Fig.~\ref{fig2}, in real space.
Read off the angle of incidence $\chi_{\mathrm{in}}$ (except for the
first iteration) and the angle of boundary orientation $\theta$. 
   \item[C.] From $\chi_{\mathrm{in}}$, find  
   the incoming $\vec{k}_{in}$, the reflected $\vec{k}_{r}$, and transmitted $\vec{k}_{t}$ wave vectors and the corresponding group velocities in momentum space as illustrated in Fig.~\ref{fig2}. Note that the $k_{\bot}$-$k_{\parallel}$ coordinate system has to be rotated by $\theta$, or alternatively, the Fermi line has to be rotated by $-\theta$, cf.~Fig.~\ref{fig3} for the example of $\theta=90^{\circ}$ with regard to the 
   orientation angle $\theta$ of the boundary. Determine the reflection angle $\chi_r$. 
   \item[D.] Calculate the reflectivity $R$ and the transmittivity $T$ by wave matching.
   \item[E.] Use $\chi_r$ to restart at B, i.e.~to find the next reflection point. 
  \end{enumerate}
  The algorithm will be explained in more detail in the following.

  \subsection{Setting initial conditions}
  The direction of the group velocity of electrons with a certain wave vector $\vec{k}$ corresponds to the normal vector on the Fermi line in this point. Due to the trigonally warped Fermi line, the emission characteristics of an electron source in bilayer graphene shows thus three preferred group velocity directions. This characteristics has to be taken into account when setting the initial conditions, see also Fig.~\ref{fig7} below. To this end, we generate the initial conditions by scanning the Fermi line evenly with respect to the polar angle $\varphi$ and calculate both the wave vector $\vec{k}$ and the corresponding group velocity $\vec{v}_{gr}$ by solving the secular equation, Eq.~\eqref{eqn:EVEq_01}, for a given Fermi energy $E_F$, the four band Hamiltonian $H^{\xi}_{BGL}$, and the $4\times4$ identity matrix $\vec{I}$ 
   \begin{equation}
    \det\left(H^{\xi}_{BLG}(\vec{k})-E_F\vec{I}\right)=0.\label{eqn:EVEq_01} 
   \end{equation}     
   Then the distribution of the initial group velocity shows the typical behavior with three preferred propagation directions.
 
  \subsection{Finding the next scattering point}
\begin{figure}[tb]
   \begin{center}
    \includegraphics[scale=2]{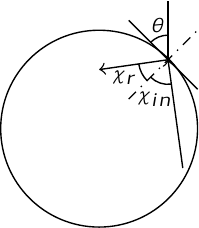}
   \end{center}
   \caption{The orientation angle $\theta$ describes the angle between the tangent to the boundary in the intersection point and the vertical taking positive values in counterclockwise direction. }
   \label{fig1}
  \end{figure}

  Starting from the last reflection point (or the initial starting point in the very first iteration) in the direction of the group velocity (that is uniquely described by the reflection angle $\chi_r$), the next intersection point with the geometric cavity boundary can be easily calculated in real space. The angle of the tangent with respect to the vertical ("$y$-axis") at this boundary point determines the orientation angle $\theta$. It is measured with respect to the vertical taking positive values in mathematically positive direction of rotation, cf.~Fig.\ref{fig1}.
 
  \subsection{Calculating the new  wave vectors}
    \begin{figure}[tb]
   \begin{center}
    \includegraphics[width=0.4\textwidth]{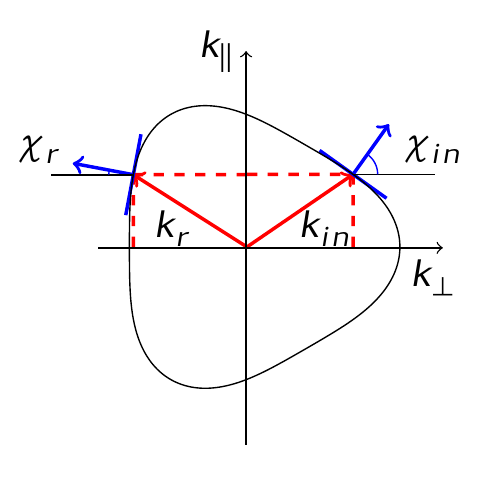}
   \end{center}
   \caption{The normal vector on the Fermi line corresponds to the group velocity. The angle between the group velocity and the x axis gives the angle of incidence or rather angle of reflection. Due to the asymmetric Fermi line, the angle of incidence is not equal to the angle of reflection.}
  \label{fig2}
  \end{figure}
  Knowing the orientation angle $\theta$ we proceed by rotating the 
  coordinate system in momentum space by $\theta$ such that the new $k_\parallel$ axis aligns parallel to the present boundary tangent.  
  Using $k_{x(y)}= p_{x(y)}/\hbar$, cf. Eq.~(\ref{eq1}, the new $k_{\bot}, k_{\parallel}$ directions then read
  \begin{align}
      k_\bot &= k_x\cos\theta-k_y\sin\theta \\
      k_\parallel &= k_x\sin\theta+k_y\cos\theta.
  \end{align}
  The wave vector components parallel to the boundary $k_{\parallel,{in}}$ 
  are conserved during the scattering process,
  \begin{equation}
   k_{\parallel,{in}}=k_{\parallel,{r}}=k_{\parallel,{t}}.
  \end{equation}
  This condition is illustrated in Fig.~\ref{fig2} and together with the corresponding Eq.~(\ref{eqn:EVEq_01}) yields the full wave vectors 
  for reflection, $\vec{k}_{r}$, when the inner Fermi line is considered, and for transmission, $\vec{k}_{t}$, for the outer Fermi line. Due to the scattering into the valence band, the transmitted electrons behave like holes and wave vector and propagation direction are opposite to each other. 
  The normal to the respective Fermi line in its intersection points with $\vec{k}_{r}$ ($\vec{k}_{t}$) yields the direction of the group velocities and hence the angle of reflection (transmission).

In addition, complex solutions $\vec{k}_{r,ev}$ and $\vec{k}_{t,ev}$ can be found which correspond to evanescent wave functions that are taken into account but 
do not contribute to the flux.


  \subsection{Wave matching}
  To this end we solve 
 Eq.~(\ref{eqn:EVEq_01}) for the eigenvectors $\Psi$ for all cases $\vec{k}_{in}$, $\vec{k}_{r}, \vec{k}_{t}, \vec{k}_{r,ev}$, and $, \vec{k}_{t,ev}$ 
 and find the 
 wave functions 
$\psi_{in}$, $\psi_{r}, \psi_{t}, \psi_{r,ev}$, and $\psi_{t,ev}$.
The matching conditions require that all wave functions have the same value at the boundary and can be used to determine 
reflection and transmission coefficients, $r$ and $t$. Setting $\Psi_{r} = r \psi_{r}$, $ \Psi_{t}=t \psi_{t}$, $\Psi_{r,ev}=r_{ev} \psi_{r,ev}$, $\Psi_{t,ev}=t_{ev} \psi_{t,ev}$, and the amplitude of the incoming wave equal to 1, $\Psi_{in} = \psi_{in}$, the wave matching condition can be written as
  \begin{equation}
   \Psi_{in}+\Psi_{r}+\Psi_{r,ev}=\Psi_{t}+\Psi_{t,ev}.\label{eqn:WMEq_01} 
  \end{equation} 

  The calculation of the transmittivity $T$ and reflectivity $R$ require the expectation value 
  of the velocity operator $v_\bot$. It is given by 
  \begin{equation}
   v_\bot=\frac{\partial H^{\xi}_{BLG}(\theta)}{\partial p_\bot}=\xi\left(\begin{matrix}
    0  & v_3\Theta & v_4\Theta^{\dagger} &v \Theta^{\dagger}\\
    v_3\Theta^{\dagger}& 0 & v\Theta &v_4\Theta\\
    v_4\Theta & v\Theta^{\dagger} & 0  & 0\\
    v\Theta & v_4\Theta^{\dagger} & 0 & 0
   \end{matrix}\right)
  \end{equation}  
  where $\Theta=\cos\theta-i\sin\theta$ and $\Theta^{\dagger}=\cos\theta+i\sin\theta$. The 
  transmittivity $T$ and reflectivity $R$ follow as
  \begin{align}
   T&=\frac{\langle\Psi_{t} | v_\bot | \Psi_{t}\rangle}{\langle\Psi_{in} | \Psi_{in}
  \rangle}\\
   R&=\frac{\langle\Psi_{r} | v_\bot | \Psi_{r}\rangle}{\langle\Psi_{in} | \Psi_{in}
 \rangle}.
  \end{align}
  
  In the case that no matching wave vector $\vec{k}_{t}$ for the transmitted wave can be found (because there is no intersection point in the outer Fermi line), total reflection with $R=1$ is assumed.

  \subsection{Outgoing trajectory}
  The reflected and transmitted trajectory characteristics are a direct result of step C. The angles of reflection and transmission, $\chi_r$ and $\chi_t$, are defined via the respective group velocities that in turn are given as the normal vector to the inner and outer Fermi line. We point out that the conventional laws, well-known from optics, for reflection and transmission are generally not fulfilled in bilayer graphene billiards.
  
  The reflection angle $\chi_r$ determines the direction of the reflected trajectory and thus the next scattering point when following the trajectory in real space. The algorithm then restarts at B. 

 \section{Further Results}
  \subsection*{Source characterisitcs and sample trajectories}
 %
  \begin{figure*}
      \centering
      \includegraphics[width=1\linewidth]{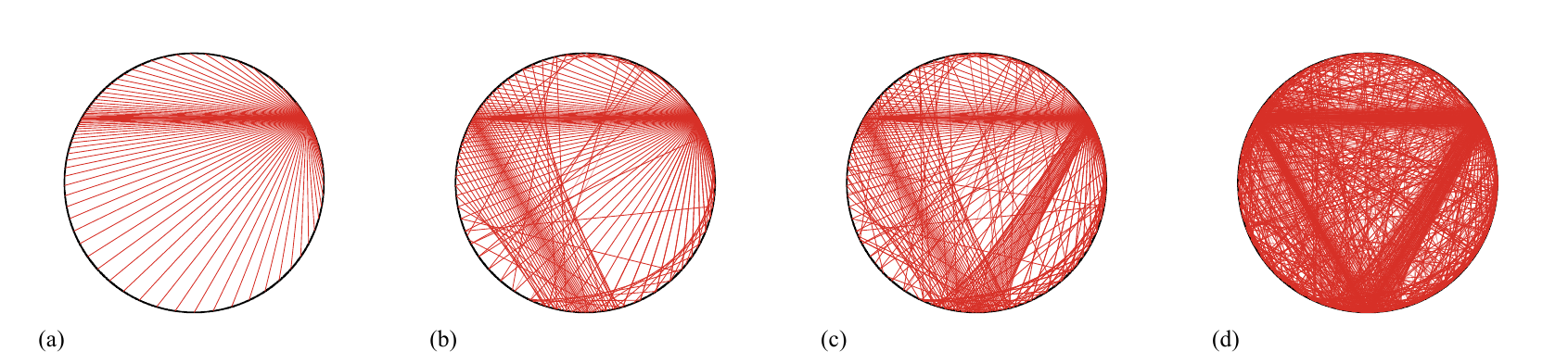} 
      \caption{Illustration of the source characteristics for a source placed at $s=1/12$,
      a stable fixed point of the triangular orbit. The source emission is anisotropic and favors directions corresponding to the flattened regions of the Fermi line. The evolution of the emitted trajectories is shown at their (a) first, (b) second, (c) third and (d) tenth reflection. The formation of the triangular orbit is clearly visible, as is the evolution of chaotic particle trajectories.}
      \label{fig7}
  \end{figure*}
 \begin{figure*}
      \centering
      \includegraphics[width=1\linewidth]{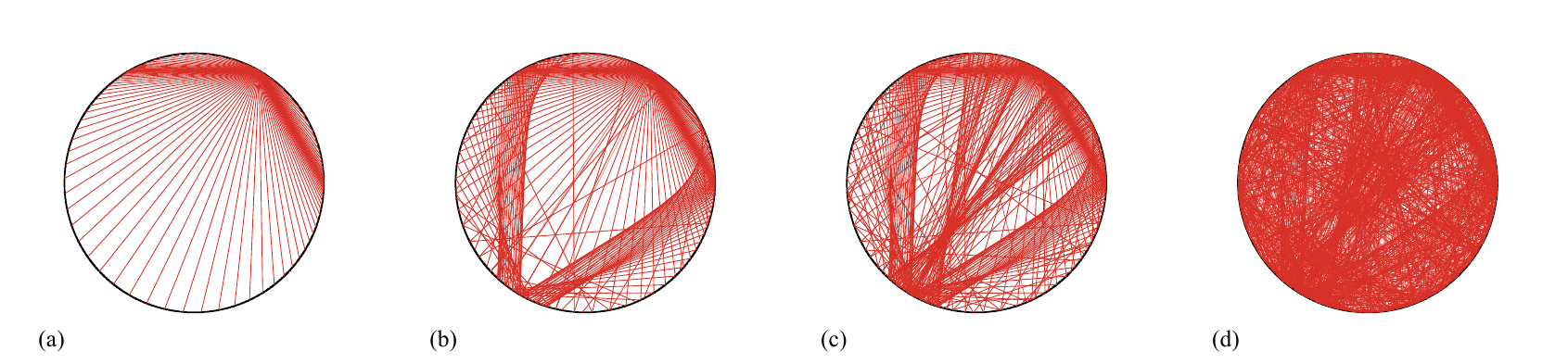} 
      \caption{Same as Fig.~\ref{fig7} but for a source placed at $s= 1/6$. No periodic orbit can be formed here as is clearly visible after several reflections in (d) where the trajectory traces homogeneously fill the whole cavity. 
      }
      
      \label{fig7_2}
  \end{figure*}
 We illustrate the billiards dynamics described by steps A-E in Figs.~\ref{fig7} and \ref{fig7_2}. Figures \ref{fig7}(a) and \ref{fig7_2}(a) show the emission from a source placed at $s=1/12$ and $s=1/6$, 
 respectively. Clearly, the source emission prefers directions that correspond to the flat regions of the Fermi line as indicated in the left part of Fig.~2 in the main text, as implied by the material anisotropy. 

  In Fig.~\ref{fig7}, the source is placed on the stable triangular orbit. Following the trajectories over the second, third, and tenth reflection in In Fig.~\ref{fig7}(b,c, and d) reveals that the triangular orbit is indeed stable but that other initial directions disturb its appearance through the chaotic trajectories started for generic initial conditions.

  Source positions outside the stable triangular orbit induce predominantly chaotic particle trajectories. This is illustrated in Fig.~\ref{fig7_2}. The source is now placed at $s=1/6$, 
  and 
  the chaotic particle dynamics dominates after a few reflections.

  \subsection*{Angles of incidence and reflection}
\begin{figure}[htb]
    \centering
           \includegraphics[width=0.4\textwidth]{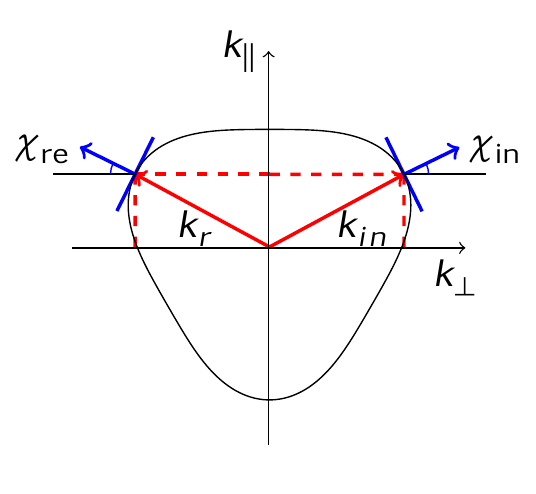}
    \caption{For certain angles $\theta$ (here $\theta=90^{\circ}$) the Fermi line is symmetric with respect to the $k_\parallel$ axis yielding $\chi_{in}=\chi_r$.}
    \label{fig3}
\end{figure}
\begin{figure}[htb]
    \centering
    \includegraphics[width=0.49\textwidth]{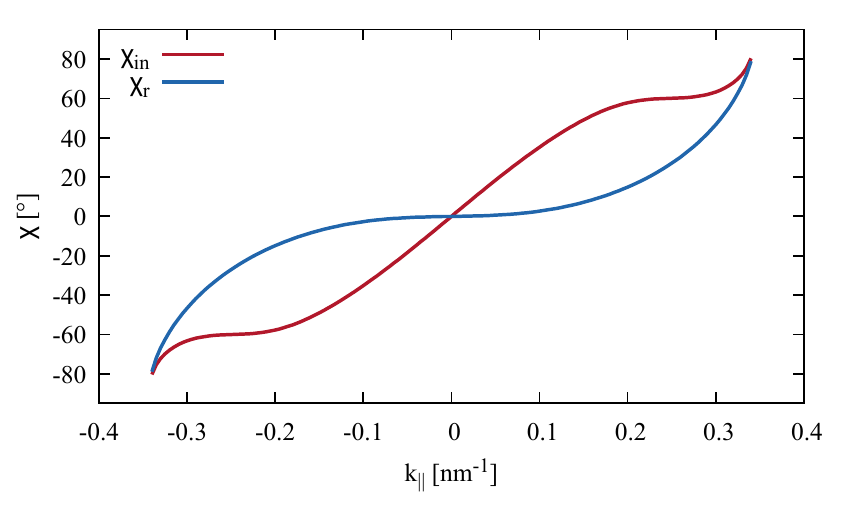}
    \includegraphics[width=0.49\textwidth]{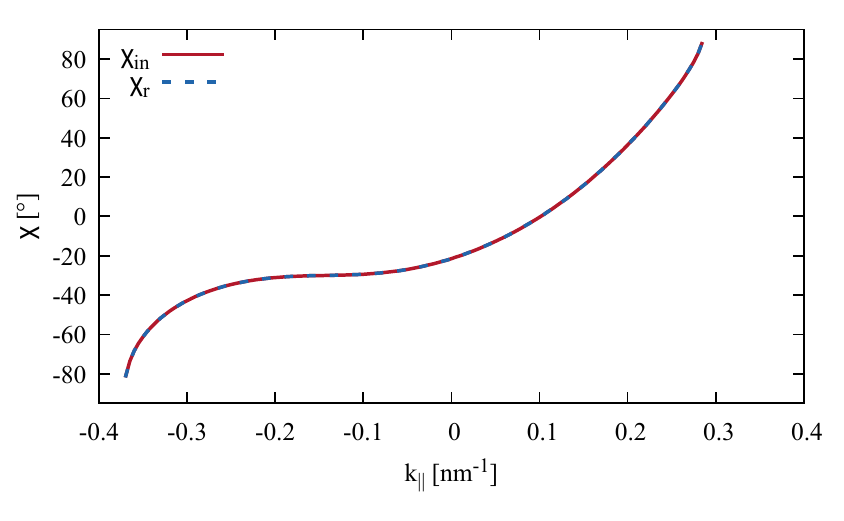}
    \caption{Connection between the angles of incidence and reflection, $\chi_{in}$ and $\chi_r$, and 
    the $y$ component $k_\parallel$ of the wave vector for a given orientation angle $\theta$.
    Whereas for $\theta=0^\circ$ (upper panel) the conventional reflection law cannot be fulfilled, it holds for $\theta=90^\circ$ (lower panel) when the Fermi line is symmetric with respect to the vertical ($k_\parallel$) axis, cf.~Fig.\ref{fig3}. 
    }
    \label{fig:angle_in_re}
\end{figure}
\begin{figure}[htb]
    \centering
   \includegraphics[width=0.45\textwidth]{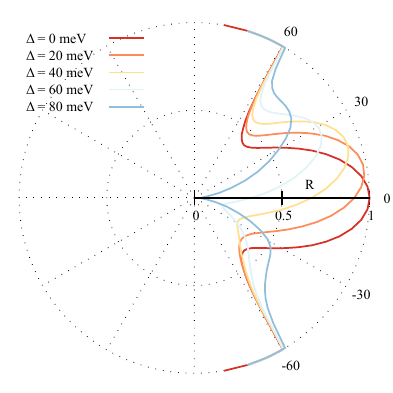}
    \includegraphics[width=0.45\textwidth]{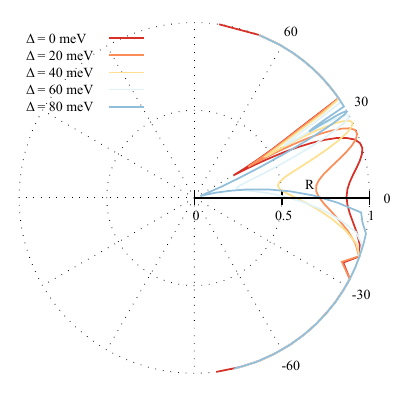}
    \caption{Polar plot of the reflection coefficient $R$ as function of angle of incidence $\chi_{in}$ for various parameters $\Delta$ and  $E_F=100$ meV, $U=145$ meV. The orientation angle $\theta$ is $\theta=0^{\circ}(30^\circ)$ in the upper (lower) panel. For $\theta=30^{\circ}$, the strong sensitivity of the reflectivity of the triangular orbit trajectory, $R(\chi_{in}=30^{\circ}$, on $\Delta$ is evident. For $\theta=0^{\circ}$, 
    anti-Klein tunneling occurs for $\chi_{in} =0^{\circ}$. }
    \label{fig:R_chi}
\end{figure}
\begin{figure*}[htb]
    \centering
    \includegraphics[width=14.5cm]{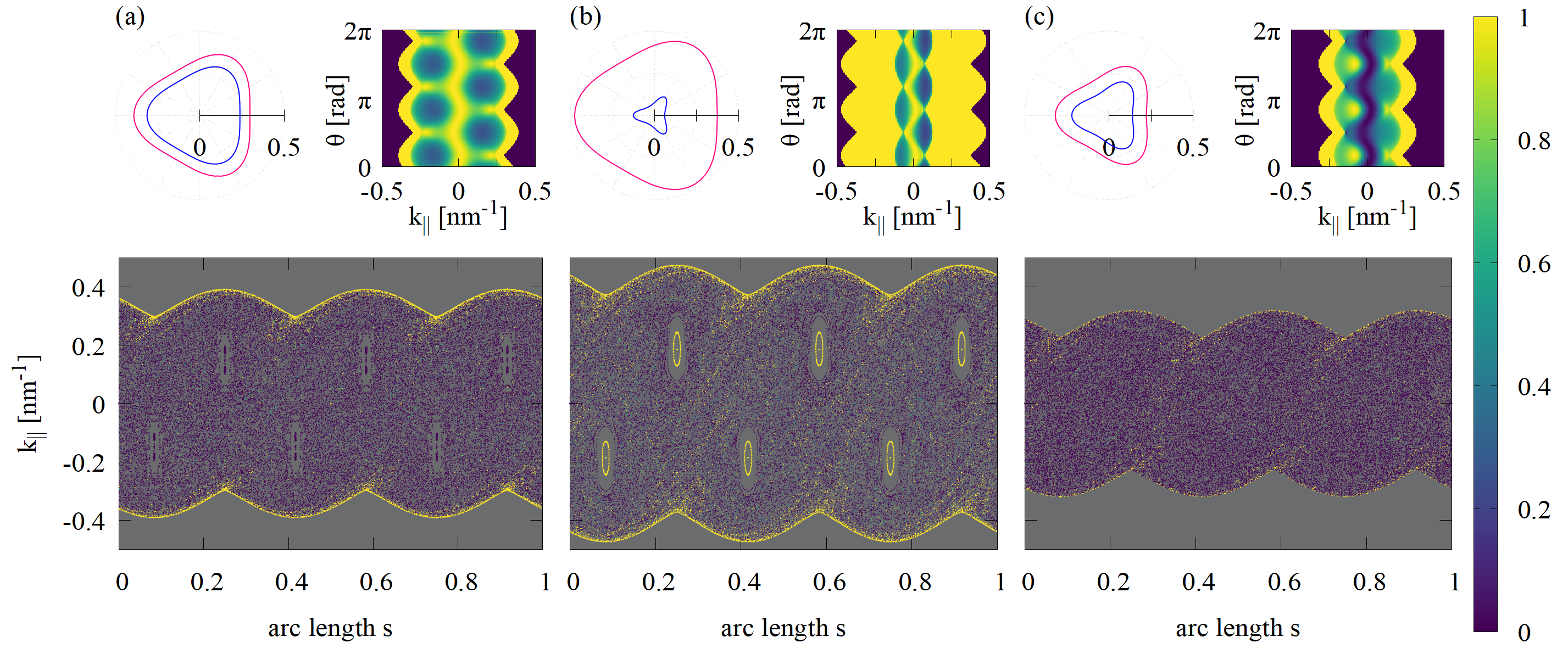}
    \caption{Same picture as Fig. 2 in the main text but for $\xi=-1$ ($K^{-1}$ valley).}
    \label{fig:xi-1}
\end{figure*}
 In the conventional, naive ray optics, reflection and transmission are governed by the law of reflection ($\chi_{in} = \chi_r$) and Snell's law. The same would hold for electron optics in case of a circular Fermi line. 

  If the Fermi line is not symmetric with respect to the $k_y$ (or $k_\parallel$) axis, the 
  group velocities of the incoming and reflected electrons (having the same $k_y$ component) cannot be mirror symmetric, cf.~Fig.~\ref{fig1}. 
  Due to the threefold symmetry of the Fermi line in the trigonally warped bilayer graphene system considered here, 
  the angle of incidence $\chi_{in}$ is usually not equal to the angle of reflection $\chi_{r}$. 
  
  However, for specific orientation angles $\theta$ the Fermi line symmetry can be restored such the the conventional  
  reflection law $\chi_{in}=\chi_{r}$ holds. This is illustrated in 
  Fig.~\ref{fig3}. Such a behavior is found 
  for $\theta$ = 
  $30^\circ,90^\circ,150^\circ,210^\circ,270^\circ,330^\circ$ and applies in particular to the stable triangular orbits discussed before, and to their unstable counterparts that arise when the stable trajectory is travelled in the opposite direction. 
  

  One consequence of this behaviour is that the conserved quantity is, in general, not the angle of incidence. Rather the momentum component $k_\parallel$ along the boundary takes this role, with a unique relation between $k_\parallel, \chi_{in }$ and $\chi_r$ that depends on the orientation angle $\theta$. This is illustrated in Fig.~\ref{fig:angle_in_re}.

\subsection*{Tuning the reflectivity $R$ via the band gap}
The reflectivity $R$ is another quantity where the rich and particular properties of bilayer graphene billiards become evident. Whereas in the optical counterpart $R$ is fixed by the refractive index ratio $n$ of the (isotropic) media, the behaviour can be easily tuned for the bilayer graphene system, for example by variation of the 
band gap $\Delta$. In Fig. \ref{fig:R_chi}, the reflectivity $R$ is shown for various $\Delta$ and two different orientation angles $\theta$ in a polar plot $R(\chi_{in})$. To this end, we have used the relation between $k_\parallel$ and $\chi_{in}$ (see Fig.\ref{fig:angle_in_re}) to express $R(k_\parallel)$ as $R(\chi_{in})$ as is usually done, e.g., in optics.

For $\theta=0^\circ$ and $\Delta=0$ the reflectivity shows symmetric behavior with respect to the angle of incidence, $R(\chi_{in}) = R(-\chi_{in})$, and anti-Klein tunneling, $R(\chi_{in}=0^\circ)=0$, cf.~the upper panel of Fig.~\ref{fig:R_chi}. This symmetry is perturbed for increasing $\Delta$ as well as for changing the orientation angle $\theta$. As a consequence, the reflectivity at a certain angle of incidence $\chi_{in}$ may vary dramatically with $\Delta$, for example near $\chi_{in}=30^\circ$ for $\theta=30^\circ$, cf.~the lower  panel of Fig.~\ref{fig:R_chi}. Note that these are precisely the parameters of the triangular orbit. Its high tunability can be used in transport measurements. 

\subsection*{Results for the other valley $K^-$}
So far we have discussed and provided the results for the 
$K^+$ valley ($\xi=1$). In Fig.~\ref{fig:xi-1}, we show the results for the other, the 
$K^-$ valley with $\xi=-1$. By changing the valley index to $\xi=-1$, the Fermi line is turned by $180^\circ$, or mirrored on the $y$ axis. Thus the results are qualitatively the same as in Fig.~2 in the main paper, with the 
Poincar\'{e} surface of section being shifted by $\Delta s/s_0=0.5$.

\end{document}